\documentclass[aps,floatfix,preprint,nofootinbib, preprintnumbers, superscriptaddress,showpacs,prd]{revtex4-1}
\usepackage[parfill]{parskip}    
\usepackage{graphicx}
\usepackage{amssymb}
\usepackage{epstopdf}
\usepackage{verbatim}
\usepackage{color}
\usepackage{url}
\usepackage{hyperref}
\usepackage[caption=false]{subfig}
\newcommand{\be}{\begin{equation}}
\newcommand{\bea}{\begin{eqnarray}}
\newcommand{\beq}[1]{\begin{equation}\label{#1}}

\newcommand{\ee}{\end{equation}}
\newcommand{\eea}{\end{eqnarray}}
\newcommand{\eeq}{\end{equation}}
\newcommand{\lsim}{\!\mathrel{\hbox{\rlap{\lower.55ex \hbox{$\sim$}} \kern-.34em \raise.4ex \hbox{$<$}}}}
\newcommand{\gsim}{\!\mathrel{\hbox{\rlap{\lower.55ex \hbox{$\sim$}} \kern-.34em \raise.4ex \hbox{$>$}}}}

\newcommand{\abs}[1]{\left| #1 \right|}

\begin{document}

\setlength{\baselineskip}{0.22in}

\preprint{MCTP-12-17}

\title{Vectorlike Fermions and Higgs Couplings}
\author{John Kearney}
\author{Aaron Pierce}
\affiliation{Michigan Center for Theoretical Physics (MCTP) \\
Department of Physics, University of Michigan, Ann Arbor, MI
48109}
\author{Neal Weiner}
\affiliation{Center for Cosmology and Particle Physics, 
Department of Physics, New York University, New York, NY 10003}
\date{\today}

\begin{abstract}
New vectorlike fermions that mix with the third generation can significantly affect the $\tau$ and $b$ Yukawa couplings.   Consistent with precision electroweak measurements, the width of the Higgs boson to  $\tau \tau$, $b \bar{b}$ can be reduced by ${\mathcal O}$(1) with respect to the Standard Model values. In the case of the $b$ quark, a reduced width would result in an enhanced branching ratio for other final states, such as $\gamma \gamma$.  New leptons can also substantially modify the Higgs boson branching ratio to photons through radiative effects, while new quarks can contribute to $gg$ fusion. The combined effect can be as much as a factor of two on the branching ratio to $\gamma \gamma$. The new quarks and leptons could be light, which would allow discovery at the LHC.  In the case of significant suppression of  $h \rightarrow \tau \tau$, searches for new leptons decaying to $\tau$-rich final states, perhaps in association with Higgs bosons are motivated.
\end{abstract}
\maketitle

\section{Introduction}
Following the observation of a ``Higgs-like" state near 125 GeV \cite{ATLASHiggs,cmshiggs}, it remains to precisely determine whether the couplings are in fact those expected from a Standard Model (SM) Higgs boson.   Branching ratios to a variety of final states are non-trivial at this mass value -- many measurements can be made and compared to their corresponding SM predictions. Deviations from these measurements could potentially indicate the presence of new physics.

If the Standard Model is a good effective theory near the weak scale, a modification to the width to fermions $\Gamma( h \rightarrow f \bar{f})$ is realized through the presence of the dimension-six operator
\begin{equation}
\label{eqn:operator}
{\mathcal O}_{h^{3}} = (f_{D} h f^{c} ) (h^\dagger h),
\end{equation}
where $f_{D}$ represents an $SU(2)$ doublet, and $f^{c}$ is the right-handed partner.
When combined with a Standard Model-like Yukawa coupling $y_f^0 f_D h f^c $, the mass and effective Higgs Yukawa coupling of the $f$ are given by
\begin{eqnarray}
\label{eqn:expandedmass}
m_{f} & = & y_{f}^{0}  v + c_{h^{3}} v^{3}, \\
\label{eqn:expandedyuk}
y_{f}^{\text{eff}} & = & y_{f}^{0}  + 3 \, c_{h^{3}} v^{2},
\end{eqnarray}
where $v = 174 \text{ GeV}$ and $c_{h^{3}}$ represents the coefficient of ${\mathcal O}_{h^{3}}$.  The mismatch between  Eqs.~(\ref{eqn:expandedmass}) and (\ref{eqn:expandedyuk}) indicates the possibility for a discrepancy between the observed fermion mass and the Yukawa coupling.  Of particular interest are modifications for $f=\tau, b$.    These couplings are small enough that it is plausible for them to be affected by integrating out new physics near the weak scale, but not so small that it is hopeless to measure them in the near future.  In this paper, we concentrate on the modification of these two couplings.

\begin{figure}
\includegraphics[width=5.75in]{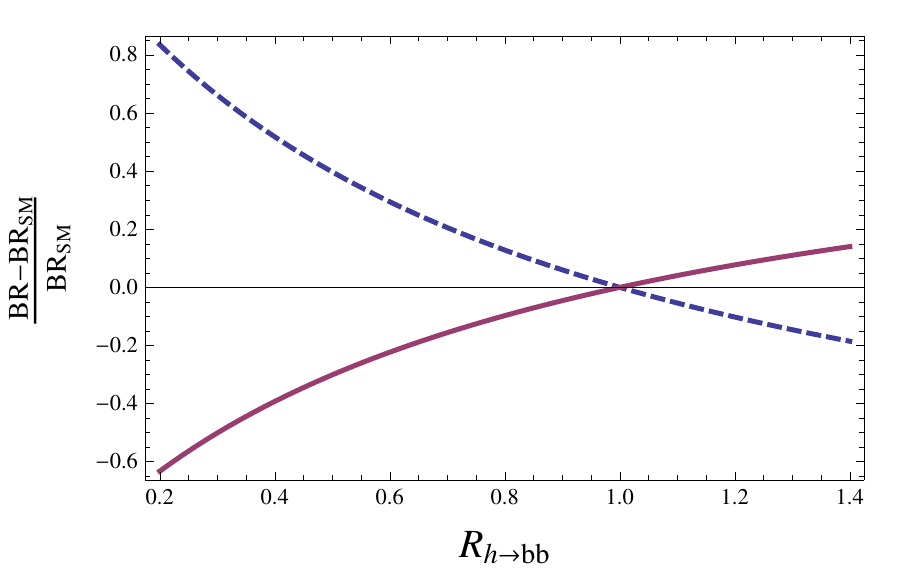}
\caption{The effect of a modification of the width of a 125.5 GeV Higgs boson to $b$ quarks (plotted as a function of $R_{h \rightarrow b b} \equiv \Gamma(h \rightarrow b \bar{b})/\Gamma_{SM}(h \rightarrow b \bar{b})$) on the branching ratio to $b$ quarks (solid) and branching ratios to all other states (dashed).  The Standard Model widths are taken from \cite{Denner:2011mq}.}
\label{fig:knockoneffect}
\end{figure}

An interesting secondary effect arises for  $f=b$.  Since for $m_{h} = 125.5$ GeV the $b$ makes up a large fraction of the total width,  a suppression (or enhancement) of $\Gamma (h \rightarrow b \bar{b})$  will affect the branching ratio (BR) to all other states.  Defining $R_{h \rightarrow b b} \equiv \Gamma(h \rightarrow b \bar{b})/\Gamma_{SM}(h \rightarrow b \bar{b})$, in Fig.~\ref{fig:knockoneffect} we illustrate the effect of a modification of the $b$ width on the BR to $b \bar{b}$ (solid) and to all other states (dashed).  For example, a 40\% reduction in $\Gamma(h \rightarrow b \bar{b})$ can give rise to a $\sim30\%$ enhancement in the branching ratios of all other channels. 

Previously, realizations of this operator have been considered in the context of two Higgs doublet models (e.g., \cite{Blum:2012kn}). Here,  we realize the operator of Eq.~(\ref{eqn:operator}) by integrating out new vectorlike fermions.  We examine how large its effect can be consistent with precision electroweak constraints and discuss its implications for Higgs phenomenology.   We highlight how these new vectorlike fermions could be of possible interest for two trends realized in the current data:  the branching ratio to $\gamma \gamma$ seems somewhat higher than expected \cite{ATLASHiggs, cmshiggs}, while the branching ratio the $\tau \tau$ is lower than expected \cite{cmshiggstau}, see, e.g., \cite{Giardino:2012dp}.  In contrast to two Higgs doublet models where, for example, the observed Higgs has minimal overlap with the down-type Higgs boson \cite{Barger:2012ky}, here the $\tau$ Yukawa may be modified independently of the $b$ Yukawa. Vectorlike leptons allow the generation of the operator of Eq.~(\ref{eqn:operator}) with $f = \tau$ without  the corresponding operator for $f=b$.   This may be of interest given the hint of a signal in $b \bar{b}$ at the Tevatron \cite{TevHiggs}.

The remainder of the paper is organized as follows.  In section \ref{sec:themodel}, we describe a model involving vectorlike fermions that can give rise to the operator of Eq.~(\ref{eqn:operator}).  We focus on two cases in which this leads to the modification of the Higgs boson coupling to $\tau$ and $b$.  In section \ref{sec:eftconstraints}, we discuss other effective field theory operators generated in this model that can be used to place constraints on the new physics.  These constraints will determine the size of the Higgs coupling modifications it is possible to achieve.  In section \ref{sec:hgg}, we discuss regions of parameter space in which these new states may also affect the effective Higgs coupling to massless gauge bosons, notably highlighting how the vectorlike leptons can sizably enhance $h \rightarrow \gamma \gamma$.  Results related to this point have recently appeared in \cite{Carena:2012xa,Joglekar:2012hb,ArkaniHamed:2012kq,Almeida:2012bq}.  We briefly explore discovery possibilities for the new states at the LHC in section \ref{sec:lhcpheno}, and comment on possible UV completions in section \ref{sec:uvcompletion}.  Finally, our conclusions are presented in section \ref{sec:conclusions}.

\section{The Model}
\label{sec:themodel}
One way to realize the  operator of Eq.~(\ref{eqn:operator}) is by mixing the Standard Model fermions with new heavy fermions. As an example, we focus first on leptons. These new leptons have additional sources of mass and, upon mixing, the $\tau$ inherits some of this mass.

We write the ``Standard Model"  tau lepton doublet as
\begin{equation}
\ell = \left(\begin{array}{c} \nu\\  \ell^-  \end{array}\right)
\end{equation}
and the corresponding $SU(2)_L$ singlet field as $e^{c}$.    We then augment the Standard Model by a vectorlike pair of $SU(2)_L$ doublets
\begin{equation}
L = \left(\begin{array}{c} N \\ L^-  \end{array}\right) \text{ and } \bar{L} = \left(\begin{array}{c} -\bar{L}^+\\  \bar{N} 
 \end{array}\right)
\end{equation}
and a vectorlike pair of $SU(2)_L$ singlets, $E^{c}$ and $\bar{E}^{c}$.   Both for simplicity and motivated by flavor constraints, we ignore mixing with the first two generations of leptons.  The presence of these fields allows several new Yukawa couplings.  The mass terms and interactions are 
\begin{equation}
- {\mathcal L} \ni y_{\tau}^{0} \ell e^{c} h + y_{E} \ell E^{c} h  + y_{L} L e^{c} h +  y_{LE} L E^{c} h + \bar{y}_{LE} \bar{L} \bar{E}^{c} h^{\dagger} + M_{E} E^{c} \bar{E}^{c}  + M_{L} L \bar{L} + \rm{h.c.}
\end{equation}
We have rotated away possible terms of the form $\mu_{\ell} \ell \bar{L}$ and $\mu_{e} e^{c} \bar{E}^{c}$ and labelled Yukawa couplings by the exotic fermion(s) present in the interaction. When the Higgs field is set to its vacuum expectation value, this leads to mass terms for the charged leptons of the form
\begin{equation}
-{\mathcal L}_{\text{mass}} = \left (\begin{array}{ccc}  e^c & E^{c} & \bar{L}^+ \end{array} \right) {\mathcal M} \left(
\begin{array}{c} \ell^- \\ \bar{E}^{c} \\ L^- \end{array} \right) + \text{ h.c.}, \text{ where } {\mathcal M} = \left(
\begin{array}{ccc}
 m^{0}_{\tau} & 0 & y_{L} v \\
y_E v & M_{E} & y_{LE} v\\
 0 & \bar{y}_{LE} v& M_{L}
\end{array}
\right)
\end{equation}
with  $v= 174$ GeV.  $\mathcal{M}$ is diagonalized to yield three charged Dirac fermions -- the $\tau$ plus two exotic, charged leptons denoted
\begin{equation}
\Psi_i = \left(\begin{array}{c} \ell_i \\ (\bar{\ell}_i)^\dagger \end{array}\right),
\end{equation}
with $i = 1,2$.  For $\mathcal{M}_{\text{D}} = \text{diag}(m_1,m_2,m_\tau) = U \mathcal{M} V^\dagger$,
\begin{equation}
\left(\begin{array}{c} \ell_1 \\ \ell_2 \\ \ell_{\tau} \end{array}\right) = V \left(\begin{array}{c} \ell^- \\ \bar{E}^c \\ L^- \end{array}\right), \text{ and } \left(\begin{array}{c} \bar{\ell}_1 \\ \bar{\ell}_2 \\ \bar{\tau} \end{array}\right) = U^\ast \left(\begin{array}{c} e^c \\ E^c \\ \bar{L}^+ \end{array}\right).
\end{equation}
The spectrum also contains the massless Standard Model neutrino and a massive neutral Dirac fermion (consisting of $N$ and $\bar{N}$) with mass $M_L$.

An analogous model can be written down to modify the effective Higgs Yukawa coupling of the $b$ by making the replacements
\begin{equation}
\ell \rightarrow q, e^c \rightarrow b^c, \stackrel{(-)}{L} \rightarrow \stackrel{(-)}{Q}, \stackrel{(-)}{E} \rightarrow \stackrel{(-)}{D}.
\end{equation}
In this case, we denote the bottom-like quarks as $B_{1}$ and $B_{2}$ respectively, and the top-like quark of mass $M_Q$ as $T$.  We neglect mixing in the top sector (e.g. via couplings of the form $\bar{Q} t^c h^\dagger$) for simplicity.\footnote{Unless the scale of new physics is very low, mixing with the SM top is unlikely to substantially modify the top Yukawa.  In any case, this could be effectively absorbed into a modification of the effective coupling of the Higgs boson to gluons and a (likely modest) modification of the Higgs boson coupling to photons.  Introduction of $\stackrel{(-)}{U}$ fields and their mixings with the $\stackrel{(-)}{Q}$ could further modify the $T$ parameter.} Since we are focused on the possible effects for Higgs physics, we assume some alignment that allows us to couple to the third generation exclusively, and do not explore flavor models explicitly.  The maximal mixing angles between the third generation and the new heavy fermions are relatively small for the benchmarks we will consider ($\sin^2 \theta \lsim 5 \times 10^{-3}$), and so given this assumption the model should be safe from flavor constraints. 

\section{Effective Theory Considerations}
\label{sec:eftconstraints}

\begin{figure}
\includegraphics[width=3.5in]{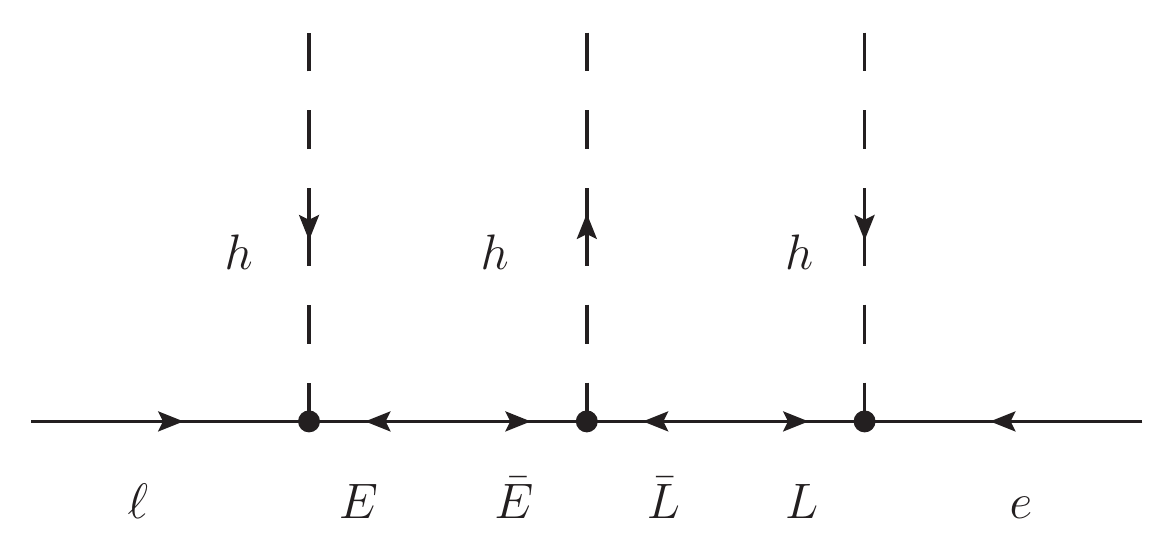}
\caption{An effective $\tau$ Yukawa coupling. }
\label{fig:eff}
\end{figure}

Integrating out the heavy lepton fields generates a contribution to the effective operator of 
Eq.~(\ref{eqn:operator}) as desired (see Fig.~\ref{fig:eff}).  The leading result is 
\begin{equation}
\label{eqn:EFTopYuk}
 c_{h^{3}} = \frac{y_{E} \bar{y}_{LE}  y_{L} }{M_{E} M_{L}}.
\end{equation}
which can take either sign, allowing it to suppress or enhance the $\tau$ Yukawa coupling relative to the Standard Model value.  We will concentrate on the suppression of Yukawa couplings, as is presently slightly preferred by the data.  Notably, this contribution to ${\mathcal O}_{h^{3}}$ is not proportional to the SM Yukawa coupling, $y_{\tau}^{0} \sim 10^{-2}$, so can potentially compete with it in spite of the $v^{2}/M^{2}$ suppression of Eq.~(\ref{eqn:expandedmass}).    Operators of yet higher dimension that include novel couplings, e.g. $y_{LE}$, can be numerically significant (and are included below), but the above equation serves as a useful guide to the expected size of the effect.\footnote{We also expect additional contributions to $c_{h^{3}}$ due to modifications of wave-function renormalization of the fermions, but these contributions are proportional to $y_{\tau}^0$ and hence negligible.}

\subsection{Constraints from the Z-pole}

The above indicates modifications of the $\tau$ and $b$ Yukawa couplings are possible, but the magnitude of the effect clearly depends on allowed size of the Yukawa couplings with the exotics.
In the case of the leptonic model, the $\bar{E}^{c}$ mixing with $\ell$ (and $\bar{L}$ mixing with $e^{c}$)  modifies the couplings of the $\tau$ lepton to gauge bosons, which can provide constraints on these couplings.

We discuss these modifications in an effective theory language \cite{BarbieriStrumia,SkibaHan} where the expressions in terms of the mass matrix are exceedingly simple.  This will serve as an important  guide to the region of parameter space where large deviations in $\Gamma(h \rightarrow f \bar{f})$ are possible, consistent with known experimental constraints.

The new physics  generates operators of the form  
\begin{eqnarray}
\label{eqn:effOpsZ1}
{\mathcal O}_{h \ell} &=& i (h^{\dagger} D_{\mu} h) (\bar{\ell}_\tau \gamma_{\mu} \ell_\tau), \\
\label{eqn:effOpsZ2}
{\mathcal O'}_{h \ell} &=& i (h^{\dagger} D_{\mu} \tau^{a} h) (\bar{\ell}_\tau \gamma_{\mu} \tau^{a} \ell_\tau), \\
\label{eqn:effOpsZ3}
{\mathcal O}_{h e} &=& i (h^{\dagger} D_{\mu} h) (\bar{\tau} \gamma_{\mu} \tau).
\end{eqnarray}
Here $\tau$ corresponds to the SU(2) singlet part of the $\tau$ lepton.
An effect of these operators is to shift the couplings to the $Z$-boson by  \cite{BarbieriStrumia,SkibaHan}
\begin{eqnarray}
\delta g_{\nu}^A = v^2  (c^{\prime}_{h\ell}-c_{h\ell}) & \qquad &  \delta g_{\nu}^V = v^2  (c^{\prime}_{h\ell}-c_{h \ell})\\
\delta g_{e}^A = v^2 (c_{he} - c_{h\ell} - c^{\prime}_{h\ell}) & \qquad & \delta g_{e}^V =  -v^2 (c_{he} + c_{h\ell}+c^{\prime}_{h \ell}).
\end{eqnarray}
Such departures in the gauge couplings are constrained by measurements at the $Z$-pole, in particular $R_{\tau} \equiv \Gamma(\rm{hadrons})/\Gamma( \tau \tau)$ as well as the asymmetries $A_{\tau}$ and $A_{FB}^{(0, \tau)}$.  The model that modifies the $b$ quark Yukawa coupling generates similar operators with $\ell_{\tau} \rightarrow q_{b}$ $\tau \rightarrow b$.  Note $R_{b} \equiv \Gamma( bb)/\Gamma(\rm{hadrons})$, the reciprocal of the definition  for the analogous $R_{\tau}$.

\begin{figure}
\includegraphics[width=5in]{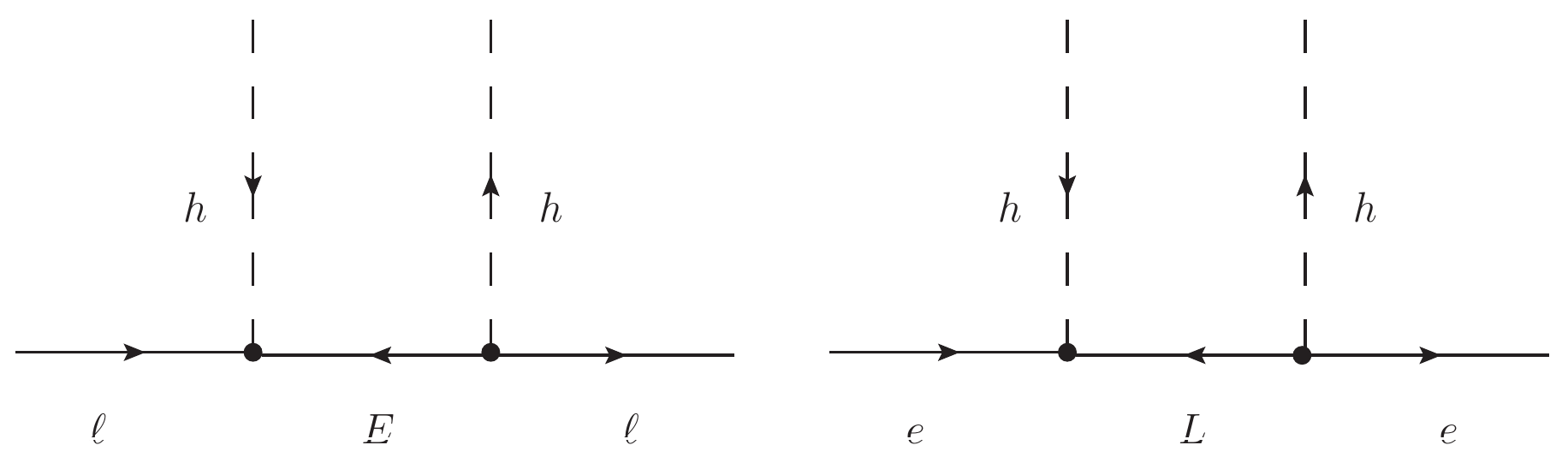}
\caption{Diagrams that give the leading contributions to the effective operators of  Eq.~(\ref{eqn:effOpsZ1})--(\ref{eqn:effOpsZ3}), and hence modification of the Z couplings to the $\tau$.} 
\label{fig:effZ}
\end{figure}

Our new physics generates ${\mathcal O}_{h \ell}$ and ${\mathcal O'}_{h \ell}$ with identical coefficients so that coupling of the neutrino (top quark) is unchanged. Consequently, the shifts in the $\tau$ couplings simplify:
\begin{equation}
\delta g_{\tau}^A = v^2 (c_{he} - 2 c_{h\ell} ), \qquad \delta g_{\tau}^V =  -v^2 (c_{he} + 2 c_{h\ell}).
\end{equation}
Constraints on these values will help determine the maximum size of the effect of Eq.~(\ref{eqn:EFTopYuk}).  The diagrams in Fig.~\ref{fig:effZ} generate $c_{he}$ and $c_{h \ell}$.  To leading order, the shifts to the $\tau$ vector and axial couplings are given by
\begin{eqnarray}
\label{eqn:eftgA}
\delta g_{\tau}^A &=& \frac{v^2}{2} \left(\frac{y_{L}^{2}}{M_{L}^2} + \frac{y_{E}^{2}}{M_{E}^{2}}\right), \\
\label{eqn:eftgV}
 \delta g_{\tau}^V &=& \frac{v^2}{2} \left(\frac{y_{L}^{2}}{M_{L}^2} - \frac{y_{E}^{2}}{M_{E}^{2}}\right).
 \end{eqnarray}
Higher order operators (e.g. operators of the form ${\mathcal O}_{h i} (H^\dagger H)^n$) can involve $y_{LE}$ and $\bar{y}_{LE}$ -- if these couplings are large, their contribution can be relevant.  As we perform exact numerical diagonalization of the relevant mass matrices, these effects are included in our discussion below.
Note that while $\delta g_{\tau}^A$ has fixed sign,  $\delta g_{\tau}^V$ can take on either sign (or be tuned small). 
The constraints on Eqs.~(\ref{eqn:eftgA}) and (\ref{eqn:eftgV}) from data limit  $y_L/M_L$ and $y_E/M_E$, and hence will limit the size of $c_{h^3}$.    With the replacements, $L \rightarrow Q$ and $E \rightarrow D$, the results of this section  trivially translate to vectorlike quarks.

\subsection{Modification of the Tau Yukawa Coupling}

We now turn towards a quantitative discussion of how measurements of $\tau$ leptons at the  $Z$ pole constrain the lepton model.  We will then be prepared to discuss the size of the modifications to the $\tau$ Yukawa coupling achievable subject to these constraints.   

To leading order in the couplings we have
\begin{equation}
\begin{array}{rclcrcl}
  R_{\tau} & \propto &(g_V^2 + g_A^2)^{-1} \qquad & \Rightarrow& \qquad \delta R_{\tau}&=& R_{\tau}^{SM} (0.3 \, \delta g^{V}_{\tau}+ 4.0 \, \delta g^{A}_{\tau}), \\
A_{\tau}  & \propto  &\frac{g_{A} g_{V}}{g_{V}^2 +g_{A}^2} \qquad &\Rightarrow& \qquad  \delta A_{\tau} &=& 0.29  \, \delta g^{A}_{\tau} -3.9 \, \delta g^{V}_{\tau}, \\
A^{0,\tau}_{FB} &=& \frac{3}{4} A_{e}^{SM} A_{\tau} \qquad &\Rightarrow& \qquad \delta A^{0,\tau}_{FB} &=& \frac{3}{4} A_e^{SM} \delta A_{\tau}.
\end{array}
\end{equation}
Experimental results for these quantities, as well as their SM predictions (using the value of $\sin^{2} \theta_{W}$  found from fitting the entire suite of PEW data) are given by \cite{PDG}:
\begin{equation}
\label{eqn:tauexpt}
\begin{array}{rclcrcl}
R_{\tau}^{exp} &=& 20.764 \pm 0.045 & \qquad R_{\tau}^{SM} &=& 20.789 \pm 0.011,\\
A_{\tau}^{exp} &=& 0.1439 \pm 0.0043 & \qquad A_{\ell}^{SM} &=& 0.1475 \pm 0.0010,\\
(A_{FB}^{(0,\tau)})^{exp}&=& 0.0188 \pm 0.0017 &  \qquad A_{FB, \tau}^{SM} &=& 0.01633 \pm 0.00021.
\end{array}
\end{equation}
The ability to make $\delta g_V^{\tau}$ small means that, in much of the parameter space, the strongest constraint comes from $R_{\tau}$.

\begin{figure}
\includegraphics[width=5.5in]{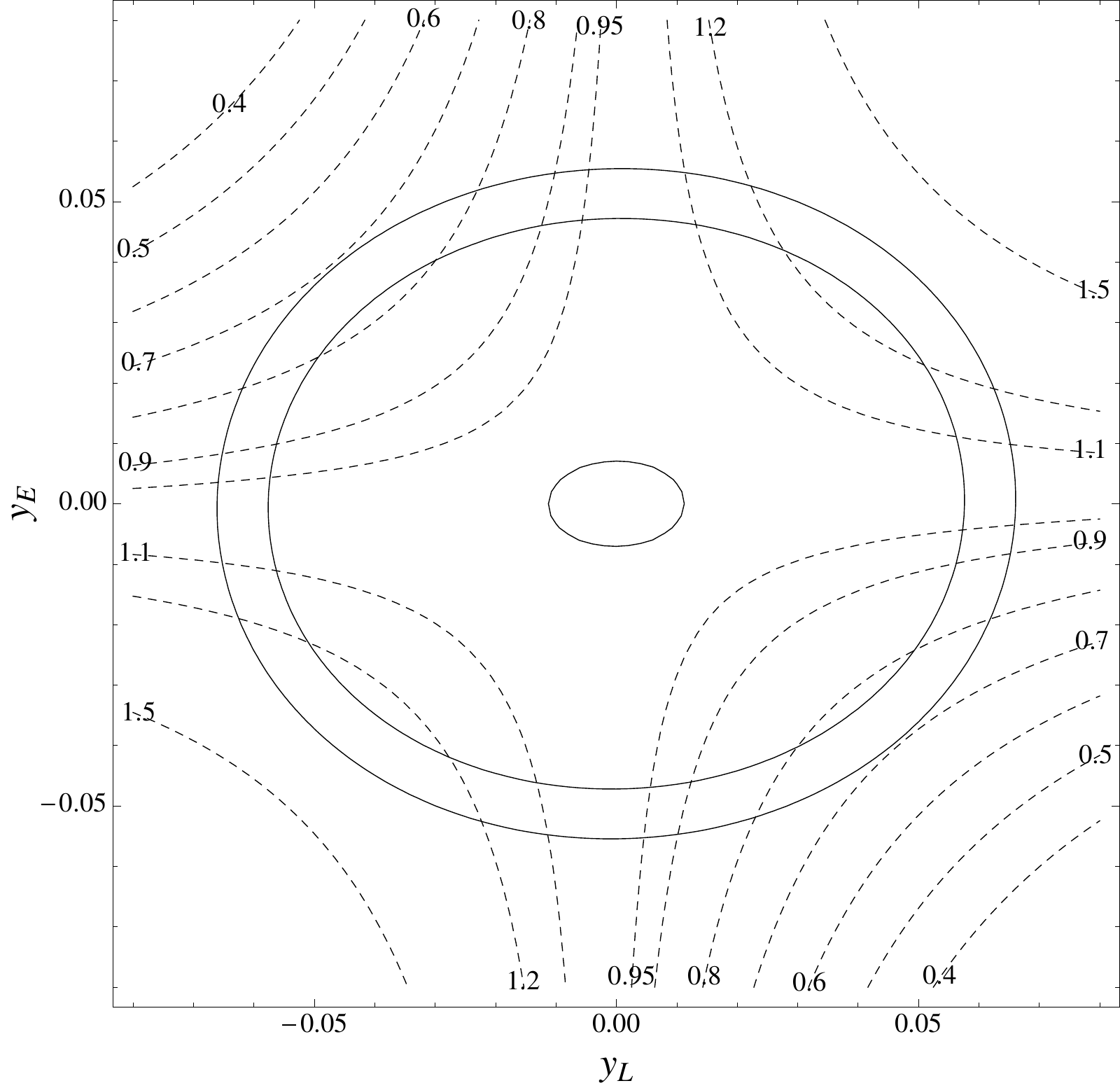}
\caption{Contours (solid) of $\Delta \chi^2 = 2.30, 4.61$, and $5.99$ relative to the best-fit value in the $(\delta g_\tau^V, \delta g_\tau^A)$ plane as a function of $y_L$ and $y_E$ for $M=M_L = M_E = 350 \text{ GeV}$ and $y_{LE} = \bar{y}_{LE} = 1$.  These values of $\Delta \chi^2$ correspond to 68.27\%, 90\% and 95\% regions for 2 parameters.  For reference, the Standard Model ($\delta g_\tau^V = 0$, $\delta g_\tau^A = 0$ or equivalently $y_L = 0$, $y_E = 0$) gives $\Delta \chi^2_{\text{SM}} = 2.09$.  Also shown are contours (dashed) of $R_{h \rightarrow \tau \tau} = \Gamma(h \rightarrow \tau \tau)/\Gamma_{\text{SM}}(h \rightarrow \tau \tau)$.  As discussed in the text, larger values of $\bar{y}_{LE}$ would allow for larger deviations in $R_{h \rightarrow \tau \tau}$.}
\label{fig:leptonbullseye}
\end{figure}

For fixed $(M_L, M_E, y_{LE}, \bar{y}_{LE})$, the constraints on $\delta g_{\tau}^V$ and $\delta g_{\tau}^A$ can be visualized in terms of elliptical $\Delta \chi^2$ contours in the  $(y_L ,y_E)$ plane.   Meanwhile,  we can see from Eq.~(\ref{eqn:EFTopYuk}) that lines of constant $R_{h \rightarrow \tau \tau} \equiv \Gamma(h \rightarrow \tau \tau)/\Gamma_{\text{SM}}(h \rightarrow \tau \tau)$ will be approximate hyperbolae in the same plane.  An example of these curves is shown in Fig.~\ref{fig:leptonbullseye} for $M_L = M_E = 350 \text{ GeV}$ and $y_{LE} = \bar{y}_{LE} = 1$.  Note  the $\Delta \chi^2$ shown is measured with respect to the global minimum in the $(\delta {g_\tau^V}, \delta g_\tau^A)$ plane.  This minimum has  $\delta g_\tau^A < 0$, which cannot be achieved in this model, i.e. no point with $\Delta \chi^{2} = 0$ appears on this plot. The largest deviation in $R_{h \rightarrow \tau \tau}$ consistent with requiring that the values of $\delta g_\tau^V$ and $\delta g_\tau^A$ give a particular $\Delta \chi^2$ can be determined by finding the hyperbola of greatest deviation that intersects the appropriate $\Delta \chi^2$ ellipse.  For instance, in the case of the reference point chosen in Fig.~\ref{fig:leptonbullseye}, one can achieve $R_{h \rightarrow \tau \tau} \approx 0.7$ consistent with a $(\delta g_\tau^V, \delta g_\tau^A)$ fit satisfying $\Delta \chi^2 < 5.99$.  Note that if one chose to allow $\Delta \chi^2$ measured instead relative to the Standard Model less than 5.99, it would not significantly alter these results, although one could reach slightly smaller values of $R_{h \rightarrow \tau \tau}$ (e.g. $R_{h \rightarrow \tau \tau} \approx 0.6$ for the reference point).

Above, we have taken $\bar{y}_{LE} \sim {\mathcal O}(1)$ in order to achieve an appreciable affect on $R_{h \rightarrow \tau \tau}$.  Even larger deviations in $R_{h \rightarrow \tau \tau}$ may be achieved by increasing $\bar{y}_{LE}$.  To a good approximation, the maximal effect on $c_{h^{3}}$ is proportional to this coupling.   Taking $\bar{y}_{LE}=y_{LE}=2$ allows $R_{h \rightarrow \tau \tau} = 0.45$. Couplings  this large can produce tension with the isospin breaking parameter $T$ that can be ameliorated by going to large $M$ (of order a TeV).  We elaborate on this issue in section~\ref{sec:Tconstraint}. We remain agnostic as to the new physics that would be required at low scales in these cases due to the presence of a Landau pole.

In contrast to $\bar{y}_{LE}$, the size of of $y_{LE}$ does not significantly affect the maximum achievable deviation in $R_{h \rightarrow \tau \tau}$.  This is especially true if it is not too large -- it contributes exclusively through higher dimension operators.  These higher dimension operators can cause the contours  of both  $\Delta \chi^2$ and $R_{h \rightarrow \tau \tau}$ to shift, but they move very little with respect to one another.  So, for a different $y_{LE}$, a somewhat different underlying choice of $y_{E}$ and $y_{L}$ may be needed to achieve a similar effect in $R_{h \rightarrow \tau \tau}$. 

The region of parameter space with both $\bar{y}_{LE}$ and $y_{LE}$ large can also potentially allow for significant enhancement of $h \rightarrow \gamma \gamma$.  We return to and elaborate on this point in section~\ref{sec:hgg}.

\subsection {Modification of the b Yukawa Coupling}
\label{sec:bYukawa}

We now turn to discuss how the measurements of $b$ quarks at the $Z$ pole constrain the quark model.  To leading order in the couplings we have
\begin{equation}
\begin{array}{rclcrcl}
  R_{b} & \propto &(g_V^2 + g_A^2) \qquad & \Rightarrow& \qquad \delta R_{b}&=& R_{b}^{SM}(1-R_{b}^{SM}) (-1.9 \, \delta g^{V}_{b}- 2.7 \, \delta g^{A}_{b}), \\
A_{b}  & \propto  &\frac{g_{A} g_{V}}{g_{V}^2 +g_{A}^2} \qquad &\Rightarrow& \qquad  \delta A_{b} &=& 0.66  \, \delta g^{A}_{\tau} -0.95\, \delta g^{V}_{\tau}, \\
A^{0,b}_{FB} &=& \frac{3}{4} A_{e}^{SM} A_{b} \qquad &\Rightarrow& \qquad \delta A^{0, b}_{FB} &=& \frac{3}{4} A_e^{SM} \delta A_{b}.
\end{array}
\end{equation}
The  factor of $(1-R_{b}^{SM})$ in the top equation comes from the modification of $\Gamma(Z \rightarrow \rm{hadrons})$ via the change in $\Gamma(Z \rightarrow b\bar{b})$.  The experimental results and SM predictions are \cite{PDG}:
\begin{equation}
\label{eqn:gconstraint}
\begin{array}{rclcrcl}
R_{b}^{exp} &=&  0.21629 \pm 0.00066  & \qquad R_{b}^{SM} &=& 0.21576 \pm 0.00004,\\
A_{b}^{exp} &=& 0.923  \pm 0.020  & \qquad A_{b}^{SM} &=&  0.9348\pm 0.0001,\\
(A_{FB}^{(0,b)})^{exp}&=&  0.0992 \pm 0.0016  &  \qquad A_{FB,b}^{SM} &=& 0.1034 \pm 0.0007.
\end{array}
\end{equation}

\begin{figure}
\includegraphics[width=5.5in]{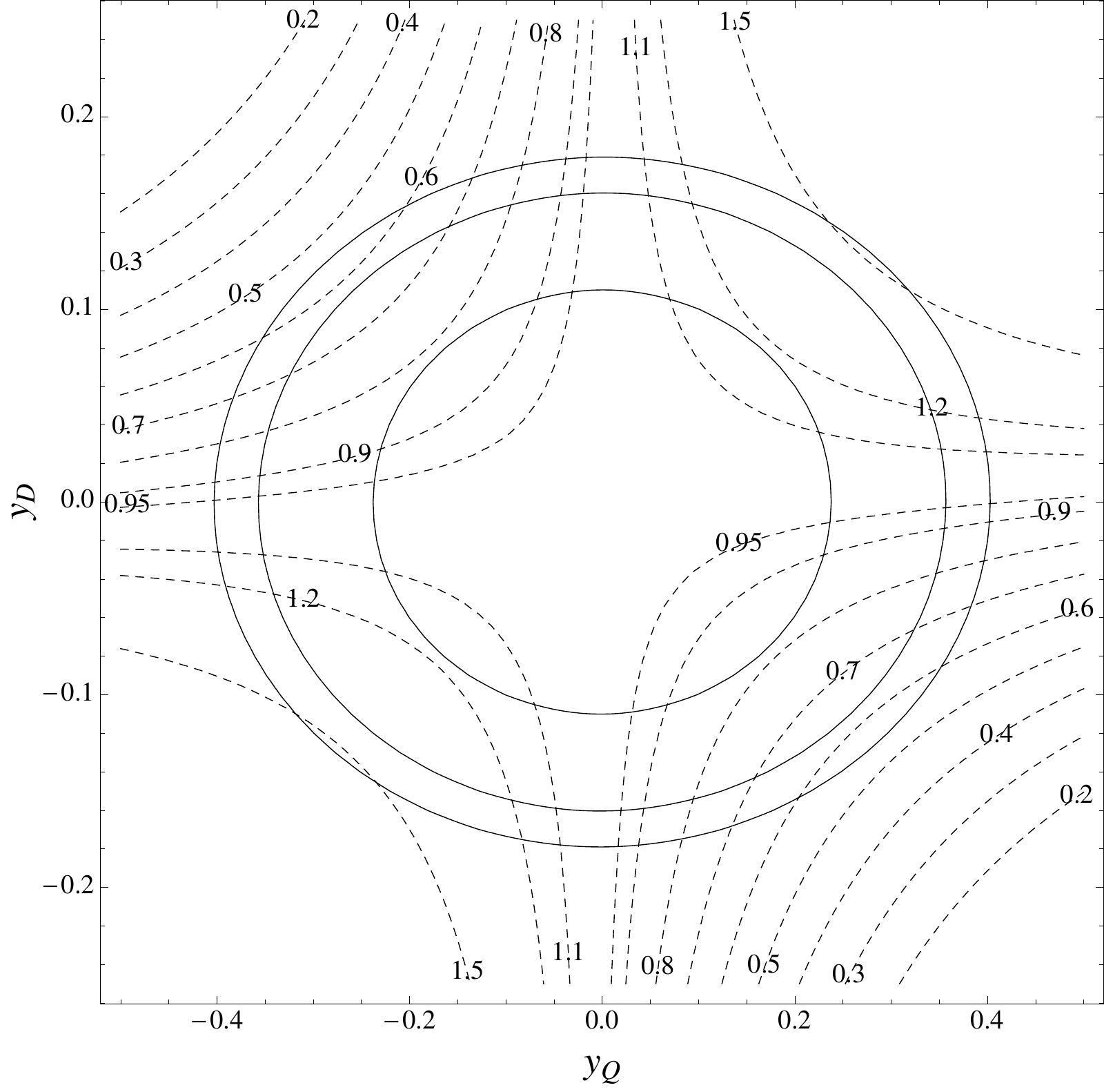}
\caption{Contours (solid) of $\Delta \chi^2 = 2.30, 4.61$, and $5.99$ relative to the best-fit value in the $(\delta g_b^V, \delta g_b^A)$ plane \emph{neglecting $A_{FB}^{(0,b)}$} as a function of $y_Q$ and $y_D$ for $M = M_Q = M_D = 600 \text{ GeV}$ and $\bar{y}_{QD} = 1$, $y_{QD} = 0$.  These values of $\Delta \chi^2$ correspond to 68.27\%, 90\% and 95\% regions for 2 parameters.  For reference, the Standard Model ($\delta g_b^V = 0$, $\delta g_b^A = 0$ or equivalently $y_Q = 0$, $y_D = 0$) gives $\Delta \chi^2_{\text{SM}} = 0.95$.  Also shown are contours (dashed) of $R_{h \rightarrow b b} = \Gamma(h \rightarrow b b)/\Gamma_{\text{SM}}(h \rightarrow b b)$.  As discussed in the text, larger values of $\bar{y}_{QD}$ would allow for larger deviations in $R_{h \rightarrow b b}$.}
\label{fig:quarkbullseye}
\end{figure}

Because $A_{FB}^{(0,b)}$ deviates from the Standard Model expectation by 2.6$\sigma$ (leading to $\Delta \chi^2_{\text{SM}} = 6.8$ with respect to the best fit  $(\delta g_b^V, \delta g_b^A)$ point), a requirement of a very small $\Delta \chi^{2}$ with respect to the global minimum in the $(\delta g_b^A, \delta g_b^V)$ plane is very difficult to satisfy.  For our purposes, we view it as an unreasonable requirement.  After all, this model is not designed to rectify this apparent discrepancy with the Standard Model (see e.g.~\cite{BeautifulMirrors}).    Instead we use the following prescription.  Neglecting $A_{FB}^{(0,b)}$ the Standard Model fit greatly improves, yielding $\Delta \chi^2_{\text{SM}} = 0.95$.  
We therefore require points exhibit small $\Delta \chi^2$ relative to the global minimum from $R_b$ and $A_b$ only -- we neglect $A_{FB}^{(0,b)}$.  We have confirmed that points with a big shift in the $b$ Yukawa do not produce a significantly worse fit to $A_{FB}^{(0,b)}$ than the Standard Model.  One can think of this $\Delta \chi^2$ for these two measurements as approximately representing the goodness of fit relative to that of the Standard Model.  As we will see, quite large modifications in the $b$ Yukawa can be achieved, so even a somewhat more stringent requirement on the $\Delta \chi^{2}$ could yield appreciable effects.

The analog of Fig.~\ref{fig:leptonbullseye} for the quark model is shown in Fig.~\ref{fig:quarkbullseye}, with $M_Q = M_D = 600 \text{ GeV}$ and $\bar{y}_{QD} = 1$, $y_{QD} = 0$ (to reduce tension with $\Delta T$ constraints -- see section~\ref{sec:Tconstraint}).  Subject to the requirement $\Delta \chi^2 < 5.99$, one can achieve $R_{h \rightarrow b b} \approx 0.55$; recalling Fig.~\ref{fig:knockoneffect}, such a modification would result in an increase of all other branching ratios by a factor of 1.34.  A more extreme choice, e.g. $\bar{y}_{QD} = \frac{3}{2}$, $y_{QD} = 0$, $\Delta \chi^2 < 5.99$, allows $R_{h \rightarrow b b} \approx 0.4$.  This point would increase branching ratios to other final states (including, interestingly, $\gamma \gamma$) by 50\%.   As promised, for these points there is not a significant degradation in the fit for $A_{FB}^{(0,b)}$: whereas the Standard Model expectation differs from the measured value of $A_{\text{FB}}^{(0,b)}$ by $2.6 \sigma$, extreme points in Fig.~\ref{fig:quarkbullseye} (those with $\Delta \chi^2 \approx 5.99$ and $R_{h \rightarrow b b} \approx 0.55$) exhibit a discrepancy at the level of $3.0 \sigma$.  

\subsection{Constraints from Oblique Corrections}
\label{sec:Tconstraint}

From comparison of Eq.~(\ref{eqn:EFTopYuk}) and Eqs.~(\ref{eqn:eftgA}-\ref{eqn:eftgV}), one can see that it is possible to avoid constraints on $\delta g^V$ and $\delta g^A$ while generating a larger $c_{h^3}$ by compensating for small $y_{E}/M_E$ and $y_{L}/M_L$ with a larger value for $\bar{y}_{LE}$. However, there are constraints on the model parameters in addition to the non-oblique corrections already discussed -- notably the new physics can induce corrections to the $S$ and $T$ parameters \cite{Peskin:1990zt,Peskin:1991sw}.

The $T$ parameter  corresponds to the effective operator
\begin{equation}
{\mathcal O}_{T} = |h^{\dagger} D_{\mu} h|^{2}, 
\end{equation}
such that $\alpha \Delta T = v^{2} c_{T}$.  This operator receives contributions of parametric size
\begin{equation}
c_{T} \sim \frac{y_{i}^{4}}{16 \pi^{2} M^{2}},
\end{equation}
for $y_i$ a Yukawa coupling to the exotic states.  In particular, there is a contribution that goes as $\bar{y}_{LE,QD}^4$.  
To reach the extreme regions of parameter space in which the deviation of $R_{h \rightarrow \tau \tau, h \rightarrow b b}$ is the greatest  
while avoiding tension with $\Delta T$, one can  retreat to higher mass scales for the vectorlike particles along with a 
corresponding increase in $y_{L,Q}$ and $y_{E,D}$.
 
 While these models also generate contributions to the $S$ parameter via the operator
\begin{equation}
{\mathcal O}_{S} = (h^{\dagger} \tau^{a} h) W^{a}_{\mu \nu} B^{\mu \nu},
\end{equation}
we find that the constraint from $S$ is typically not significant.  In fact, due to the correlation (88\%) between these parameters in the precision electroweak fit \cite{PDG}, a small positive contribution to $S$ typically allows somewhat larger $T$ values.

\begin{figure}
\includegraphics[width=6in]{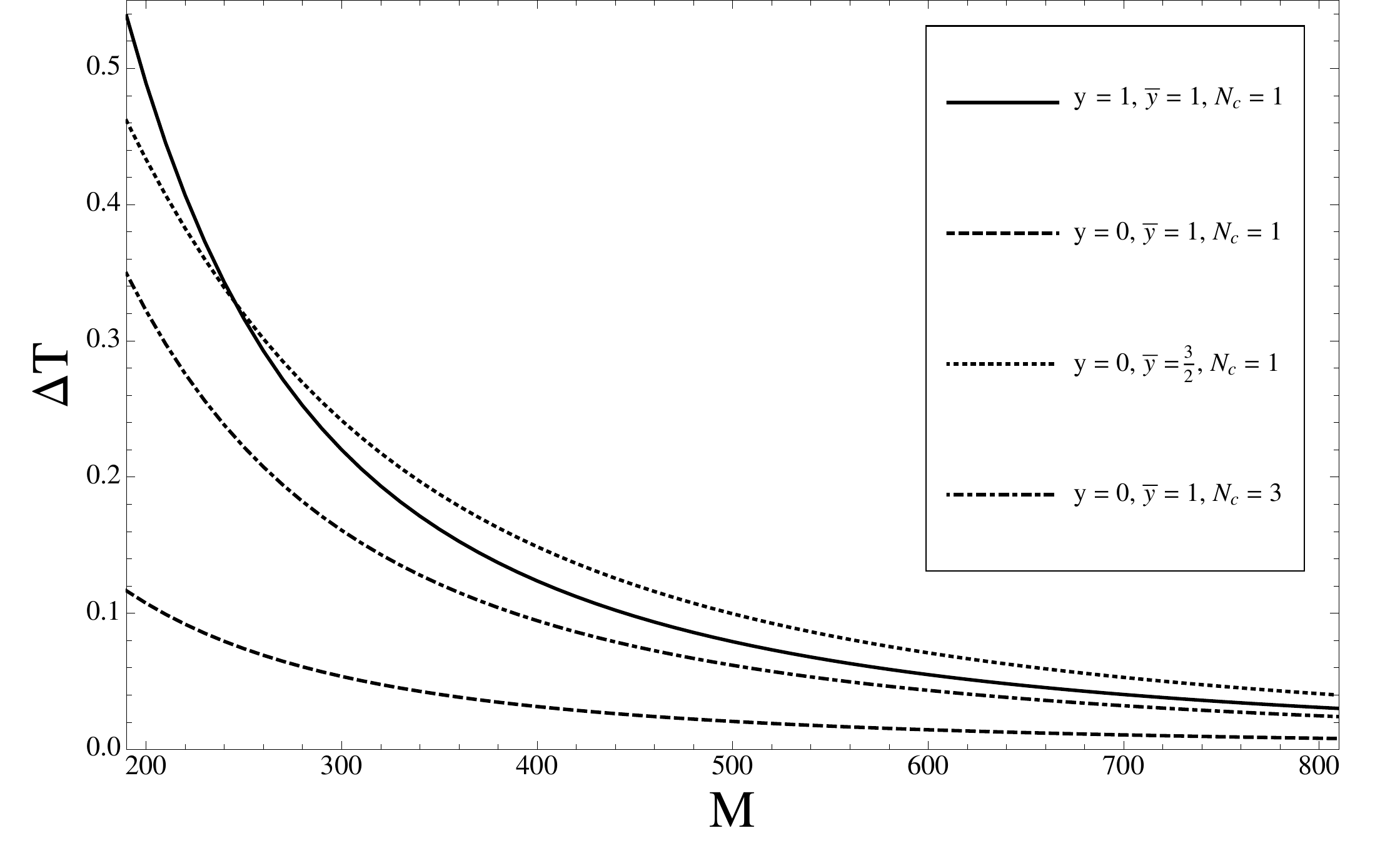}
\caption{Curves of $\Delta T$ for models with either a vectorlike lepton ($N_{c}=1$) or a vectorlike quark ($N_{c}=3$) and particular choices of the Yukawa couplings between the heavy fermions. In all cases we have taken the vectorlike doublet mass equal to the vectorlike singlet mass.}
\label{fig:DeltaT}
\end{figure}

The experimental bounds on $\Delta T,\Delta S$ at  2$\sigma$ (making a very slight adjustment for $m_{h}=125.5$ GeV) are \cite{PDG} 
\begin{eqnarray}
 -0.08 &< \Delta T <& 0.23 ,\\
-0.14 &< \Delta S  <& 0.22. 
\end{eqnarray}
Values for $\Delta T$ are shown in Fig.~\ref{fig:DeltaT} as a function of $M = M_{L,Q} = M_{E,D}$ for several choices of $\bar{y} = \bar{y}_{LE,QD}$ and $y = y_{LE,QD}$, neglecting mixing with SM particles (which is constrained to be small).  Full expressions for the contribution to $T$ from vectorlike particles can be found in \cite{Lavoura:1992np} and are reproduced for this specific model in Appendix~\ref{app:rhoparameter}.   Formulae for $S$ can also be found by suitable modification of the formulae in \cite{Lavoura:1992np}, but the expressions are more lengthy.

For $y = \bar{y} = 1$ ($N_c = 1$), the $\Delta T$ constraint requires $M \gsim 300 \text{ GeV}$ at $2 \sigma$ ($\Delta T = 0.22$, $\Delta S=0.09$) and $M \gsim 360 \text{ GeV}$ at $1 \sigma$ ($\Delta T = 0.15$, $\Delta S=0.06$).  As can be seen from the figure ($y = 0$, $\bar{y} = 1$, $N_c = 1$), one can abate the tension with $\Delta T$ and permit lower mass scales by taking $y$ small -- this suppresses contributions to $c_T$ involving $y$, and leads to smaller values for $\Delta T$.\footnote{In fact, for some negative choices of $y$ it is possible to achieve some cancellation between contributions to $T$ -- we do not concentrate on this region of parameter space as it would lead to a suppression in the $\gamma \gamma$ rate, which at present is disfavored by the data.}  For instance, one could take $y = 0$ and increase $\bar{y}$ to $\bar{y} = \frac{3}{2}$ -- in the lepton case ($N_c = 1$), this would allow $R_{h \rightarrow \tau \tau} \approx 0.6$ consistent with $\Delta \chi^2 < 5.99$ (compared with $R_{h \rightarrow \tau \tau} \approx 0.7$ for $y_{LE} = \bar{y}_{LE} = 1$) while  increasing the required mass scale to $M \gsim 380 \text{ GeV}$ (corresponding to a lightest exotic charged lepton mass of  270 GeV).
For $y_{LE} = 0$, $\bar{y}_{LE} = 2$,  $R_{h \rightarrow \tau \tau} \approx 0.5$ is allowed, with required masses $M \gsim 700 \text{ GeV}$.  The lightest lepton in this case will be at 550 GeV, likely out of reach for the LHC.  In both of these cases the contribution to $S$ is modest, and the $\Delta T$ nearly saturates the relevant bound of $\Delta T \lsim 0.16$.

In the case of the quark model, tension with constraints on $\Delta T$ is increased due to the additional color factor $N_c = 3$.  Thus, as in section~\ref{sec:bYukawa}, we consider the region of parameter space with $\bar{y}_{QD} \sim {\mathcal O}(1)$ and $y_{QD}$ small to avoid $\Delta T$ constraints.  Furthermore, as we explain in section~\ref{sec:hgg}, the region $y_{QD} \sim \bar{y}_{QD} \sim {\mathcal O}(1)$ is less interesting for $h \rightarrow \gamma \gamma$ for the quark model than for the lepton model, in part due to the reduced charge.  Consequently, increasing $y_{QD}$ would mostly serve to increase bounds on $M_Q, M_D$.  

Again, one can take larger values for $\bar{y}_{QD}$ to achieve smaller $R_{h \rightarrow b b}$ at the price of increasing bounds on $M$.  As mentioned earlier, $\bar{y}_{QD} = \frac{3}{2}$, $y_{QD} = 0$ admits $R_{h \rightarrow b b} \sim 0.4$.  The price is that the $\Delta T$ constraint requires $M \gsim 680 \text{ GeV}$.  Nevertheless, the lightest quark mass in this case is 560 GeV, likely discoverable soon. For $\bar{y}_{QD} = 2$, $y_{QD} = 0$, $R_{h \rightarrow b b}$ can be as low as $\sim 0.25$, but the mass bound increases to $M_Q = M_D = M \gsim 1250 \text{ GeV}$.  Such modifications would correspond to increases of other all branching ratios by factors of 1.5 and 1.75, respectively.

To reiterate, if one is willing to permit Landau poles at a relatively low scale, one can go to large values of $\bar{y}_{LE,QD}$ and correspondingly larger values of $M_{L,Q}, M_{E,D}$ and $y_{L,Q}$, $y_{E,D}$ to achieve smaller values for $R_{h \rightarrow \tau \tau, h \rightarrow b b}$ (or, equivalently, larger $c_{h^3}$).  However, this requires the new states to be more massive, making direct discovery more difficult, particularly in the case of the vectorlike leptons.  


\section{Modification of Higgs Couplings to Massless Gauge Bosons}
\label{sec:hgg}

\subsection{Coupling to Photons}
An extension similar to the one considered here  was discussed in \cite{Carena:2012xa} with an eye toward increasing the $\gamma \gamma$ branching ratio for the Higgs (and more recently in \cite{Joglekar:2012hb,ArkaniHamed:2012kq,Almeida:2012bq}).  There, vectorlike leptons were added without substantial (in some cases any) mixing with Standard Model leptons.

In our case, we have shown  (by dialing the $\tau$-exotic mixing) that sizable deviations to Higgs-$\tau$ effective Yukawa can be achieved for new leptons of essentially any exotic mass.  It is therefore straightforward  to simultaneously enhance the effective Higgs coupling to photons by focusing on the low mass region.  In effective theory language, we are generating the operator 
\begin{equation} 
{\mathcal O}_{\gamma \gamma}= h^{\dagger} h F^{\mu \nu} F_{\mu \nu},
\end{equation}
with coefficient  $c_{\gamma \gamma}$.
Parametrically, we expect contributions  
\begin{equation}
c_{\gamma \gamma} \sim \frac{e^{2} y_{i}^{2}}{ 16 \pi^{2} M^{2}},
\end{equation}
with $y_{i}$ the largest Yukawa coupling in the problem.  In particular, since $y_{L}$ and $y_{E}$ are small to satisfy constraints on the $Z-{\tau}-{\tau}$ coupling, we expect dominant contributions from $y_{i} = y_{LE}$ or $\bar{y}_{LE}$.   In fact, a simple expression for the coefficient of the operator that couples a single Higgs boson to photons
\begin{equation}
{\mathcal O}_{h \gamma \gamma} =  \frac{\alpha}{16 \pi}\frac{h}{\sqrt{2} v} F^{\mu \nu} F_{\mu \nu},
\end{equation}
can be derived via the general formula, \cite{Carena:2012xa}
\begin{equation}
c_{h \gamma \gamma} = b_{1/2} \frac{\partial}{\partial \log v} \log\left(\det {\mathcal M^{\dagger}_{f} } {\mathcal M_{f} } \right),
\end{equation}
with $b_{1/2} = (4/3) N_{c} Q_{f}^{2}$ for a Dirac fermion.  Using our mass matrix, neglecting the small mixing with the Standard Model leptons, we find to leading order in $v^2/M^2$
\begin{equation}
c_{h \gamma \gamma} = - \frac{16}{3} N_c Q_f^2 \frac{y_{LE} \bar{y}_{LE} v^2}{M_E M_L}
\end{equation}
with $Q_{L} = 1, N_c = 1$ for leptons. 
Maximizing constructive interference with the $WW$ loop (destructive interference with the top fermion loop) requires $y_{LE}$ and $\bar{y}_{LE}$ of same sign and large.  This further motivates our choice of $y_{LE} = \bar{y}_{LE} = 1$ in the plots of Fig.~\ref{fig:leptonbullseye}.

Armed with the above effective field theory understanding, we have identified the region of parameter space that gives the maximal change in $h \rightarrow \gamma \gamma$.  We now proceed to a full numerical evaluation of the effects.  The modification of the Higgs boson width to photons can be written as:
\begin{eqnarray}
\frac{\Gamma(h \rightarrow \gamma \gamma)}{\Gamma_{\text{SM}}(h \rightarrow \gamma \gamma)} &=&
\frac{\left| \frac{ g_{h WW}}{M_{W}^2}  A_1(\tau_{W}) + \frac{2 g_{h t \bar{t}}}{m_t} \, 3 \, \left(\frac{2}{3}\right)^{2} A_{1/2}(\tau_{top}) +\sum_{\ell_i} \frac{2 g_{h \ell_i \bar{\ell}_i}}{m_{\ell_i}}  A_{1/2}(\tau_{\ell_i}^{i}) \right|^2}
{ \left|\frac{g_{h WW}}{M_{W}^2}  A_1(\tau_{W}) + \frac{2 g_{h t \bar{t}}}{m_t} \, 3 \, \left(\frac{2}{3}\right)^{2} A_{1/2}(\tau_{top})\right|^{2}}\\
&=& \left|1- 0.109 \sum_{\ell_{i}} \frac{2 g_{h \ell_i \bar{\ell}_i} v}{m_{\ell_i}} A_{1/2}(\tau_{\ell_i}^{i}) \right|^{2}
\end{eqnarray}
with $\tau_{i} \equiv 4 m_{i}^{2}/m_{h}^{2}$, and loop integrals $A_{i}(\tau)$ defined in the appendix.
In the second line, for the particles with all of their mass from EWSB, we have substituted $\frac{g_{hWW}}{M_W^2} = \frac{2 g_{h t \bar{t}}}{m_t} = \frac{\sqrt{2}}{v}$ ($v = 174 \text{ GeV}$), as well as the values for the SM loops, $A_{W} = -8.34$, $A_{top}=1.38$.

\begin{figure}
\vspace{-,25in}
\includegraphics[width=4.5in]{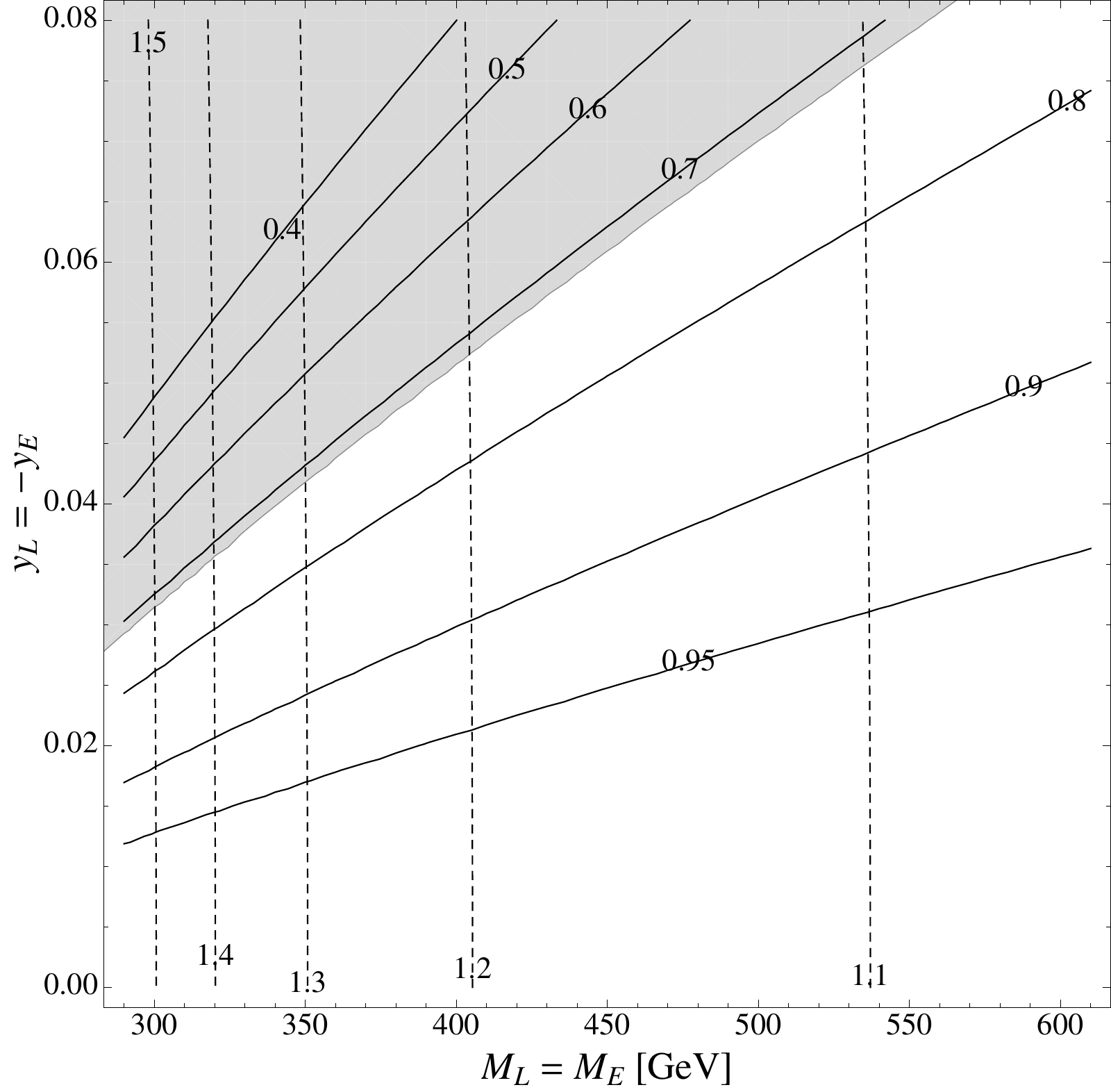}
\vspace{-.1in}
\caption{Contours of $R_{h \rightarrow \gamma \gamma} \equiv \frac{\Gamma(h \rightarrow \gamma \gamma)}{\Gamma_{\text{SM}}(h \rightarrow \gamma \gamma)}$ (dashed) and $R_{h \rightarrow \tau \tau}$ (solid) as a function of the mixing $y_{E} = -y_{L}$ and $M_{L} = M_{E}$.
We fix $\bar{y}_{LE}=y_{LE}=1$, which is allowed by precision electroweak measurements for values of $M > 300$ GeV at $2 \sigma$.  The shaded region is excluded by too large shifts $\delta g_\tau^V, \delta g_\tau^A$ in the coupling of the $\tau$ to the $Z$ ($\Delta \chi^2 > 5.99$).}
\label{fig:MasterPlot}
\end{figure}

$R_{h \rightarrow \tau \tau}$ and $R_{h \rightarrow \gamma \gamma} \equiv \Gamma(h \rightarrow \gamma \gamma)/\Gamma_{\text{SM}}(h \rightarrow \gamma \gamma)$ are shown in Fig.~\ref{fig:MasterPlot} for $\bar{y}_{LE} = y_{LE} = 1$ as a function of $M_L = M_E$ and $y_L = - y_E$.    Also shown are regions excluded by constraints on $\delta g_\tau^{V,A}$. The choice $y_L = - y_E$ is motivated by Fig.~\ref{fig:leptonbullseye}, which indicates that this is the region of parameter space that exhibits the largest deviation in $R_{h \rightarrow \tau \tau}$.  As can be seen from the figure, it is possible to simultaneously achieve $R_{h \rightarrow \tau \tau} \approx 0.7$ and $R_{h \rightarrow \gamma \gamma} \approx 1.5$ consistent with experimental constraints.  Doing so requires light new states, making it conceivable that they may be observed at the LHC -- we briefly discuss potential signatures of these light leptonic states in section~\ref{sec:lhcpheno}.

Throughout, for simplicity, we have taken the vectorlike doublet and singlet to have a common mass scale, $M_L = M_E = M$.  It is reasonable to wonder how sensitive our conclusions are to this choice.  $R_{h \rightarrow \tau \tau}$ and $\delta g_\tau^{V,A}$ are (to leading order) functions of $y_i/M_i$ ($i = L, E$), so the values they attain are largely unaffected by perturbations from this point -- shifts in $M_L, M_E$ can be compensated by corresponding shifts in $y_L, y_E$.  The situation is slightly different for the $T$ parameter: as $\Delta T$ is a measure of mass splitting within the doublet, it is more sensitive to $M_L$ than $M_E$.  Thus, it is possible to achieve the same values of $\Delta T$ by decreasing $M_E$ and increasing $M_L$ by a smaller amount (such that the splitting in the doublet decreases).  Such movements in parameter space could be used to slightly increase $R_{h \rightarrow \gamma \gamma}$ without diminishing the electroweak fit.  However, doing so only increases $R_{h \rightarrow \gamma \gamma}$ by ${\mathcal O}(5\%)$, so we consider points with $M_L = M_E$ to be appropriately representative of the variations in $R_{h \rightarrow \gamma \gamma}$ achievable.

For similar parameters, the contributions of down-type quarks to $R_{h \rightarrow \gamma \gamma}$ are  less important by a factor of three.  Furthermore, their $M_{Q,D}$ are constrained by collider (and $T$ parameter) considerations to be larger than the lepton case.  For instance, for $M_Q = M_D = M$ and $\bar{y}_QD \sim y_{QD} \sim {\mathcal O}(1)$, the lightest new state has mass $m_1 \sim M - v$.  Current bounds on vectorlike quarks constrain $M \gsim 600 \text{ GeV}$ in this case \cite{Chatrchyan:2012yea,CMS:2012ab,Rao:2012gf} -- for such values of $M$, deviations of $R_{h \rightarrow \gamma \gamma}$ from unity are negligible.
Consequently, we find that loop contributions to $h \rightarrow \gamma \gamma$ from the vectorlike quarks are generally small.  A non-trivial enhancement in the Higgs branching ratio to photons can still be achieved as a result of the suppression of the effective $b$ Yukawa.  This approach requires that the Tevatron excess in $b \bar{b}$ was not due to the Higgs boson.  Hopefully, searches for $hZ$ with a boosted $h \rightarrow b \bar{b}$ at the LHC will soon help shed light on this point.

\subsection{Coupling to Gluons}

While the vectorlike quarks do not significantly affect $h \rightarrow \gamma \gamma$, they may affect the Higgs coupling to gluons through generation of the effective operator
\begin{equation} 
{\mathcal O}_{g g}= h^{\dagger} h G^{a \mu \nu} G^a_{\mu \nu}.
\end{equation}
As in the case of photons, we can consider the coefficient of the operator that couples a single Higgs boson to gluons
\begin{equation}
{\mathcal O}_{h gg} =  \frac{\alpha_s}{16 \pi}\frac{h}{\sqrt{2} v} G^{a \mu \nu} G^a_{\mu \nu},
\end{equation}
which is generated with coefficient \cite{PhysRevD.41.1001}
\begin{equation}
c_{h g g} = b^s_{1/2} \frac{\partial}{\partial \log v} \log\left(\det {\mathcal M^{\dagger}_{f} } {\mathcal M_{f} } \right)
\end{equation}
where $b_{1/2}^s = (2/3)$ for a Dirac fermion.  To leading order in $v^2/M^2$ (and neglecting small mixing with the Standard Model), the vectorlike quarks generate a coefficient of size
\begin{equation}
c_{h g g} = - \frac{8}{3} \frac{y_{QD} \bar{y}_{QD} v^2}{M_D M_Q}.
\end{equation}
Thus, depending on the relative signs of $y_{QD}$ and $\bar{y}_{QD}$, the contribution from the new vectorlike quarks can interfere either constructively or destructively with the top loop (which dominates the Standard Model contribution).  In full, the modification of the Higgs boson gluon fusion production cross section is 
\begin{eqnarray}
\frac{\sigma(g g \rightarrow h)}{\sigma_{\text{SM}}(g g \rightarrow h)} &=&
\frac{\left|\frac{2 g_{h t \bar{t}}}{m_t} A_{1/2}(\tau_{top}) +\sum_{B_i} \frac{2 g_{h B_i \bar{B}_i}}{m_{B_i}}  A_{1/2}(\tau_{B_i}^{i}) \right|^2}
{ \left|\frac{2 g_{h t \bar{t}}}{m_t} A_{1/2}(\tau_{top})\right|^{2}}\\
&=& \left|1 + 0.512 \sum_{B_i} \frac{2 g_{h B_i \bar{B}_i} v}{m_{B_i}}  A_{1/2}(\tau_{B_i}^{i}) \right|^{2}.
\end{eqnarray}

Recent studies \cite{Carmi:2012in, Giardino:2012dp, Buckley:2012em} have suggested that fits to the data may favor a slight decrease in Higgs production via gluon fusion.  However, these studies were performed without the recent data from ATLAS on $WW$ \cite{HWWATLAS}, which favor a slightly increased rate.  
In this model, moderate enhancement or suppression of $g g \rightarrow h$ are both possible.  For $y_{QD} \sim 0$, effects on $g g \rightarrow h$ will be small, so a large deviation in $R_{h \rightarrow b b}$ from unity can be achieved without simultaneously affecting the Higgs boson production cross section.
Destructive interference with the top loop occurs for same sign $y_{QD}$ and $\bar{y}_{QD}$.  For $y_{QD} \sim \bar{y}_{QD} \sim {\mathcal O}(1)$ and $M_Q = M_D \gsim 600 \text{ GeV}$ (to avoid direct search bounds for the lightest state), one can achieve $R_{g g \rightarrow h} \equiv \sigma(g g \rightarrow h)/\sigma_{\text{SM}}(g g \rightarrow h) \approx 0.7$ (simultaneous with $R_{h \rightarrow b b} \approx 0.8$).  
Alternatively, one can generate $R_{g g \rightarrow h} > 1$ for opposite sign $y_{QD}$ and $\bar{y}_{QD}$.  For $\bar{y}_{QD} \sim - y_{QD} \sim {\mathcal O}(1)$ and $M_Q = M_D \gsim 560 \text{ GeV}$ (to satisfy bounds on a charge-$2/3$ quark decaying exclusively to $bW$ \cite{CMS:2012ab}), it is possible to achieve $R_{g g \rightarrow h} \approx 1.35$.

\section{Phenomenology}
\label{sec:lhcpheno}

\begin{figure}
\vspace{-0.7in}
\includegraphics[width=0.9\textwidth]{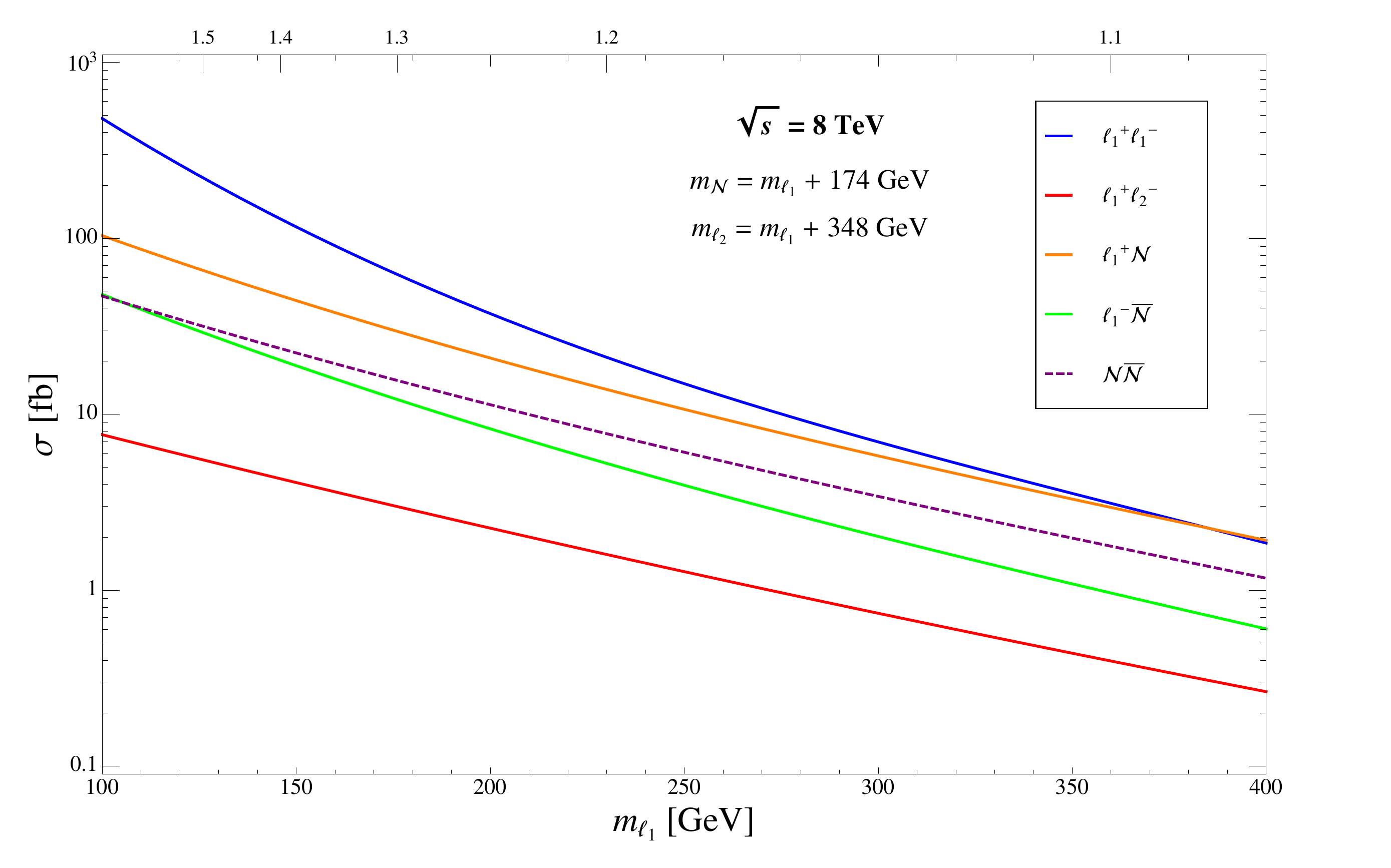}
\includegraphics[width=0.9\textwidth]{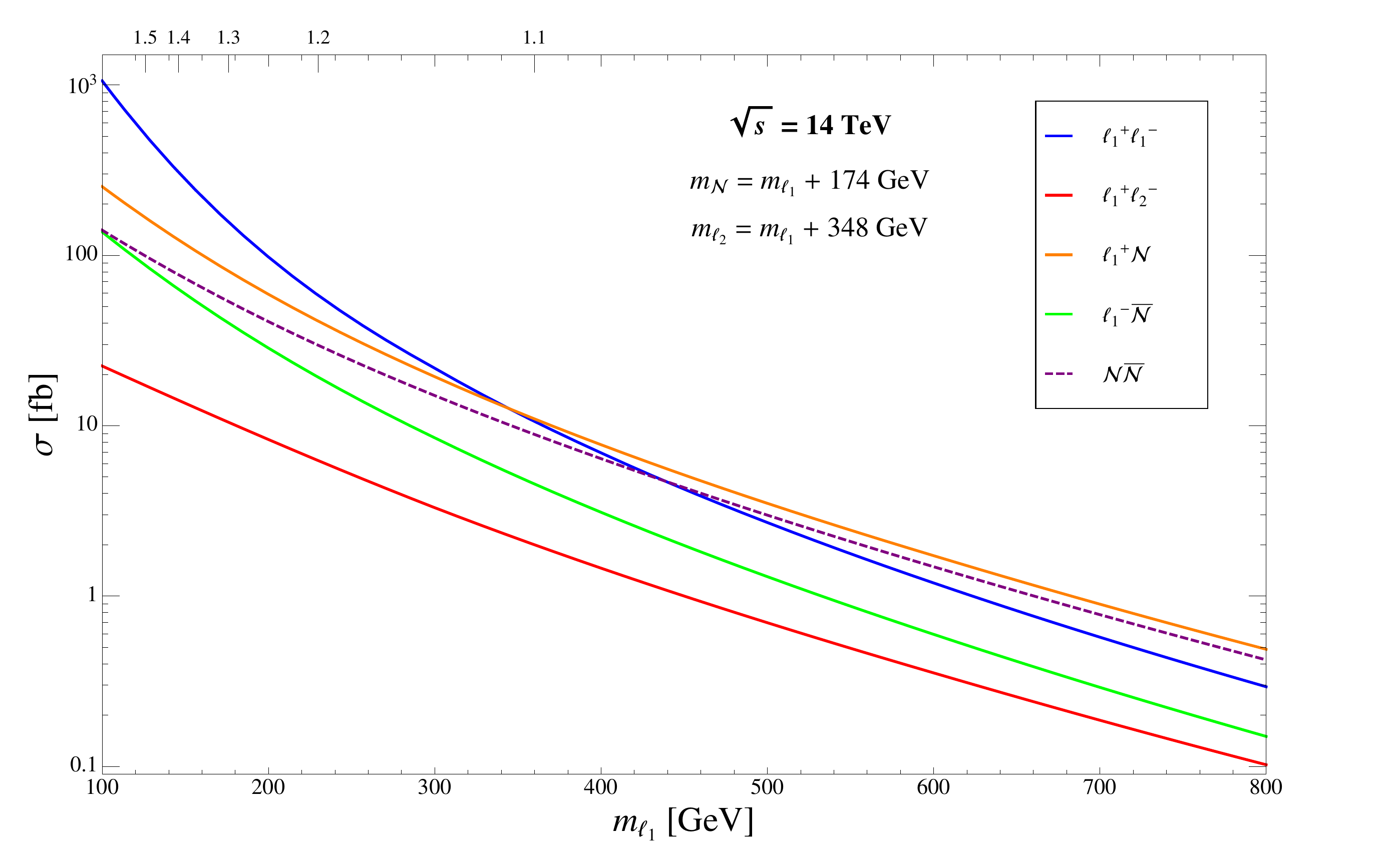}
\caption{Production cross sections for the heavy lepton states for $\bar{y}_{LE} = y_{LE} = 1$ and $M_{L}= M_{E}$ at the LHC with $\sqrt{s}$ = 8 TeV (top) and 14 TeV (bottom) as a function of the lightest exotic lepton mass.  Along the top axis the enhancement in the $\gamma \gamma$ branching ratio with respect to the Standard Model value is shown.  Note for $\bar{y}_{LE} = y_{LE} = 1$, constraints on $(\Delta T, \Delta S)$ require $M_{L} = M_E > 300 \text{ GeV}$, or $m_{\ell_1} > 126 \text{ GeV}$.  These plots are valid for any values of $y_L, y_E$ consistent with constraints on $\delta g^V_\tau, \delta g^A_\tau$.}
\label{fig:productionLHC}
\end{figure}

We have implemented the lepton model in MadGraph 5 \cite{Madgraph} using FeynRules \cite{FeynRules}.
Production rates for the new heavy leptons are shown in Fig.~\ref{fig:productionLHC}, using the CTEQ 6l1 parton distribution function.
For the points we have chosen ($y_{LE} = \bar{y}_{LE} = 1, M_{E} = M_{L}$) the masses obey the simple relation $m_{\ell_{1}} + 174 \rm{\; GeV} = m_{N} = m_{\ell_{2}} - 174 \rm{\; GeV}$.  Bounds on this model exist from LEP near the kinematic limit ($m_{{\ell_{1}}} >$ 105 GeV) \cite{LEPChargino}.

In general, the lightest charged state, $\ell_{1}$, will decay to either $\tau h$, $\tau Z$ or $W \nu$.  One potentially relevant search  is the ATLAS slepton/chargino dilepton search \cite{ATLASslepton}.  But while it overlaps the final state, it is not yet sensitive, although future searches might be at low masses.  Notable is the $\tau h$ final state \cite{Martin:2012dg}, which is particularly important in this region of parameter space.   The presence of both a vectorlike $L, \bar{L}$ and $E,\bar{E}$ allows deviation from the characteristic 1:1:2 ratio of final states found for models with only a singlet vectorlike partner (i.e. a ``Littlest Higgs"-like model).  For example, for points with $y_{LE} = \bar{y}_{LE}=1$ (as shown in the plot) and $y_{E} = -y_{L}$ large (so as to give a large suppression in $R_{h \rightarrow \tau \tau}$), the branching ratio $BR(\ell_{1} \rightarrow h \tau)$ is near 50\%.  Of course, for this final state to be relevant, the lepton must be  heavy enough to evade the phase space suppression, which limits its production cross section.  Nevertheless, we find the BR is nearly 50\% already at $m_{\ell_{1}}=175$ GeV.  With high luminosity, one might even explore the possibility of utilizing $h \rightarrow \gamma \gamma$ decays along the lines of \cite{Azatov:2012rj}, particularly given the possible enhancement of the $\gamma \gamma$ rate. The $N$ state will almost exclusively decay to $\ell_{1} W$, raising the possibility of multi-lepton cascades.

In the case of vectorlike quarks, the phenomenology is similar to well-explored heavy quark models --  
production is dominated by QCD processes, and cross sections can be determined as a function of mass \cite{Aliev:2010zk, cmsBtobZ}.
We review some of the relevant limits here (see also \cite{Okada:2012gy}).  CMS  has searched for a $B^{\prime}$ via $B^{\prime} \rightarrow b Z$.  For a $B^{\prime}$ with a 100\% BR to $bZ$, the limit is $m_{B^{\prime}} > 550$ GeV \cite{cmsBtobZ}.   A similar search from ATLAS, but using less data, sets a limit $m_{B^{\prime}} > 400$ GeV \cite{:2012ak}.  Using 4.9 fb$^{-1}$ of data the CMS collaboration excludes a  $B^{\prime}$ decaying to $tW$ with 100\% BR below 611 GeV at 95\% confidence \cite{Chatrchyan:2012yea}.  

Searches for $T \rightarrow t Z$  by CMS \cite{Chatrchyan:2011ay} exclude a top partner with 100\% BR to $tZ$ at 95\% CL.  Searches for $bW$ final states (with two leptonically decaying $W$'s) \cite{CMS:2012ab} exclude masses up to 557 GeV.  A search in the semileptonic final state by ATLAS \cite{ATLAS:2012aw} has a more limited reach of 480 GeV.  A combination and reinterpretation of these searches can be used to bound a $T$ with non-trivial branching ratios to all of $tZ$, $bW$ and $th$ \cite{Rao:2012gf}.    Some degradation of the above limits exists, perhaps by up to 100 GeV or so.  A dedicated search for $th$ would improve the situation.  In our case, depending on the exact implementation of the top sector, the $T$ will likely first cascade to a $B_{1}W$, followed by a further decay to SM fermions and electroweak bosons.    This opens the possibility of, e.g., $WWZZbb$ final states.  For $\bar{y}_{QD} \sim - y_{QD} \sim {\mathcal O}(1)$ (such that gluon fusion $g g \rightarrow h$ is enhanced), $T$ can be the lightest new state and will consequently decay to $tZ$, $bW$ and $th$ -- neglecting mixing in the top sector, $T$ decays exclusively to $bW$ and the bound of $m_T > 557 \text{ GeV}$ applies.
More detailed collider studies are left for further work.

\section{Comments on UV Completions}
\label{sec:uvcompletion}
One possibility is that the model presented here might be embedded in a (perhaps somewhat split \cite{SplitNima,SplitGian,SplitJames}) SUSY scenario, where the scalars superpartners are sufficiently heavy that they do not affect the phenomenology discussed here.  In this case, approximate gauge coupling unification is maintained only via the introduction of complete $SU(5)$ multiplets.  With a single 
$5 + \bar{5}$ and $10+ \overline{10}$ perturbative gauge coupling can be maintained. So, introduction of the vectorlike leptons would also motivate the presence of the vectorlike quarks -- this would be good news for LHC phenomenology.  In our analyses in the previous sections, we have not adhered too strictly to this motivation, as it would imply relationships between Yukawa couplings in the (exotic) lepton and quark sectors that we have not imposed.
 The ${\mathcal O}$(TeV) masses considered here could be explained by whatever mechanism is responsible for the solution to the $\mu$ term (e.g. Giudice--Masiero or NMSSM like physics).

\begin{figure}
\includegraphics[width=4in]{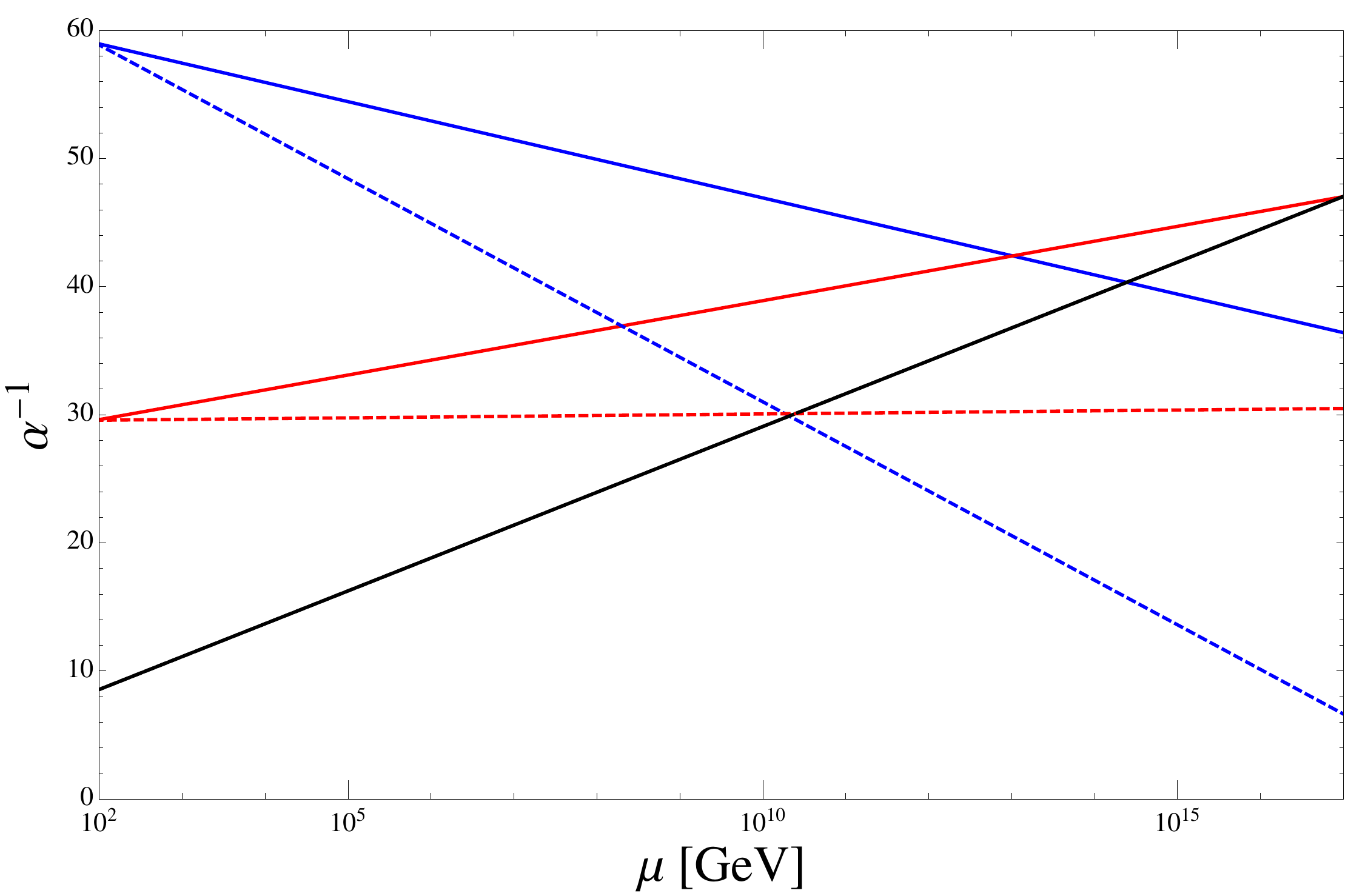}
\caption{The dashed lines indicate the one-loop running of the gauge couplings in the extension considered here with three pairs of vectorlike leptons.  The solid lines indicate the running in the Standard Model.}
\label{fig:UnificationFun}
\end{figure}

It is amusing to note the possibility of a less conventional unification story.  Suppose only vectorlike leptons are added (without the corresponding quarks).  With three pairs (i.e., $3 \times( L, \bar{L}, E, \bar{E}$)), as might be expected if there is a vectorlike partner for each generation, there is adequate unification, see Fig.~\ref{fig:UnificationFun}.
Unification can be assessed by examining the ratio $R\equiv(b_3 - b_2)/(b_2-b_1)$, where $b_{i}$ denotes the $\beta$-function for the gauge group $i$.  Under the assumption of unification, at  one loop $R = (\alpha^{-1}_{3} - \alpha^{-1}_{2})/(\alpha^{-1}_{2} - \alpha^{-1}_{1})$.   Experimentally, the RHS evaluated at $M_{Z}$  yields $0.718 \pm 0.003$.  In the Standard Model, $R=.528$, indicating an exceedingly poor fit for unification.  Augmenting by three vectorlike pairs, we find $R=.707$, which while could  plausibly be accounted for by additional corrections at the GUT scale.  The scale of unification is quite low $M_{U} = 2.4 \times 10^{10}$ GeV, so additional GUT model building (likely with some type of extra-dimensional unification) would be necessary to avoid too large dimension six proton decay.  However, it should be noted that this approach to unification creates new doublet-triplet splitting problems.  In addition, for larger Yukawa couplings (where the largest effects on the Higgs branching ratios are achieved), there is a danger of introducing vacuum instability with decay times shorter than the observed lifetime of the universe \cite{ArkaniHamed:2012kq}.

\section{Conclusions}
\label{sec:conclusions}

We have explored the possibility that vectorlike fermions can mix with the Standard Model, giving substantial modification to the Higgs boson properties.  To summarize the types of deviations possible, we have constructed Table \ref{tab:benchmarks}, which illustrates the extent to which $R_{h \rightarrow \tau \tau}$, $R_{h \rightarrow b b}$ and $R_{h \rightarrow \gamma \gamma}$ can be modified.  While the points in this table represent some of the more extreme cases (even allowing a total enhancement of $R_{h \rightarrow \gamma \gamma} \approx 2$), it should be noted that there are any number of effects in play here that can work in concert, all quite plausibly present if there is new vector like matter with significant Yukawa couplings at the TeV scale.  

\begin{table}

\setlength{\tabcolsep}{4pt}

\subfloat[$R_{h \rightarrow \tau \tau} = 0.53$]{
\begin{tabular}{| c | c | c | c | c || c | c | c || c | c | c |} \hline
$M_L = M_E$ [GeV] & $\bar{y}_{LE}$ & $y_{LE}$ & $y_L$ & $y_E$ & $\Delta T$ & $\Delta S$ & $\Delta \chi_\tau^2$ & $R_{h \rightarrow \tau \tau}$ & $R_{h \rightarrow \gamma \gamma}$ & Masses [GeV] \\ \hline
700 & 2.0 & 0.0 & 0.105 & $-0.105$ & 0.16 & 0.01 & 5.98 & 0.53 & 1.00 & 547, 700, 895 \\ \hline
\end{tabular}}

\subfloat[$R_{h \rightarrow \tau \tau} = 0.74$ and $R_{h \rightarrow \gamma \gamma} = 1.50$]{
\begin{tabular}{| c | c | c | c | c || c | c | c || c | c | c |} \hline
$M_L = M_E$ [GeV] & $\bar{y}_{LE}$ & $y_{LE}$ & $y_L$ & $y_E$ & $\Delta T$ & $\Delta S$ & $\Delta \chi_\tau^2$ & $R_{h \rightarrow \tau \tau}$ & $R_{h \rightarrow \gamma \gamma}$ & Masses [GeV] \\ \hline
300 & 1.0 & 1.0 & 0.032 & $-0.028$ & 0.22 & 0.09 & 5.90 & 0.74 & 1.50 & 126, 300, 474 \\ \hline
\end{tabular}}

\subfloat[$R_{h \rightarrow b b} = 0.25$]{
\begin{tabular}{| c | c | c | c | c || c | c | c || c | c | c |} \hline
$M_Q = M_D$ [GeV] & $\bar{y}_{QD}$ & $y_{QD}$ & $y_Q$ & $y_D$ & $\Delta T$ & $\Delta S$ & $\Delta \chi_b^2$ & $R_{h \rightarrow b b}$ & $r^{BR}_{\text{non}-b}$ & Masses [GeV] \\ \hline
1250 & 2.0 & 0.0 & 0.57 & $-0.27$ & 0.16 & 0.02 & 5.93 & 0.25 & 1.74 & 1091, 1250, 1437 \\ \hline
\end{tabular}}

\subfloat[$\frac{\text{BR}(h \rightarrow \gamma \gamma)}{\text{BR}_{\text{SM}}(h \rightarrow \gamma \gamma)} = 1.99$]{
\begin{tabular}{| c | c | c | c | c || c | c | c || c | c | c |} \hline
$M_Q = M_D$ [GeV] & $\bar{y}_{QD}$ & $y_{QD}$ & $y_Q$ & $y_D$ & $\Delta T$ & $\Delta S$ & $\Delta \chi_b^2$ & $R_{h \rightarrow b b}$ & $r^{BR}_{\text{non-}b}$ & Masses [GeV] \\ \hline
1500 & 1.5 & 0.0 & 0.71 & $-0.31$ & 0.04 & 0.01 & 5.86 & 0.39 & 1.53 & 1378, 1500, 1638 \\ \hline \hline
$M_L = M_E$ [GeV] & $\bar{y}_{LE}$ & $y_{LE}$ & $y_L$ & $y_E$ & $\Delta T$ & $\Delta S$ & $\Delta \chi_\tau^2$ & $R_{h \rightarrow \tau \tau}$ & $R_{h \rightarrow \gamma \gamma}^{\tau}$ & Masses [GeV] \\ \hline
350 & 1.0 & 1.0 & 0.0 & 0.0 & 0.16 & 0.06 & $\Delta \chi^2_{\text{SM}}$ & 1.00 & 1.30 & 176, 350, 524 \\ \hline
\end{tabular}}

\caption{Benchmark points exhibiting (a) large deviation in $R_{h \rightarrow \tau \tau}$ from unity (lepton sector), (b) moderate deviation in $R_{h \rightarrow \tau \tau}$ from unity and maximal value of $R_{h \rightarrow \gamma \gamma}$ from the lepton sector, (c) large deviation in $R_{h \rightarrow b b}$ from unity (quark sector) and (d) large enhancement of $h \rightarrow \gamma \gamma$ due to combination of additional loop contributions from vectorlike leptons and suppression of $b$ Yukawa from mixing with vectorlike quarks.  $r^{BR}_{\text{non}-b} \equiv BR(h \rightarrow X)/BR_{\text{SM}}(h \rightarrow X)$ for $X \neq b \bar{b}$ as a result of the decrease in the $h \rightarrow b \bar{b}$ width.  Note that, for points (a) and (c), we have taken $M$ as small as possible consistent with EWPT constraints -- for larger values of $M$, the same $R_{h \rightarrow \tau \tau,h \rightarrow b b}$ could be achieved with smaller values of $\Delta T$, $\Delta S$.  For point (d), $R_{h \rightarrow \gamma \gamma}^\tau$ denotes the enhancement of $h \rightarrow \gamma \gamma$ from the lepton sector only.}
\label{tab:benchmarks}
\end{table}

For example, the observed $\sigma \times BR$ for photons can be affected in multiple ways:  vectorlike quarks might enhance the gluon production cross section, the presence of vectorlike leptons might increase $\Gamma(h \rightarrow \gamma \gamma)$, and mixing with the $b$ quark ($\tau$ lepton) could reduce $\Gamma(h \rightarrow b \bar{b})$ $(\Gamma(h \rightarrow \tau \bar{\tau})$). These effects can be sizeable: we have found realizations where $\Gamma_{h \rightarrow \tau \bar{\tau}}$ is suppressed by a factor of 2, where $\Gamma_{h \rightarrow b \bar{b}}$ is suppressed by a factor of 4, where $\Gamma_{h \rightarrow \gamma \gamma}$ is enhanced by 50\% and where $\sigma_{g g \rightarrow h}$ is enhanced by 35\%, all consistent with precision constraints. Some of these effects push against the same experimental limits, so while some variations can be thought of independently, others (such as simultaneous enhancements to $\sigma_{g g \rightarrow h}$ and $\Gamma_{h \rightarrow \gamma \gamma}$) cannot be. Nonetheless, one can find points where the overall inclusive signal of $h \rightarrow \gamma \gamma$ is enhanced by a factor of 2. 

It is important to emphasize that all these effects are naturally present in models with vectorlike fermions. One need not focus on pushing every mode to the limit to produce an interesting effect. Even if individually modest, multiple contributions can combine to give an interesting effect. A combination of $\sim 10-20\%$ effects to yield a $50\%$ modification is easily plausible without pushing the limits of existing constraints.

All but the most extreme  deviations in the fermion couplings discussed here will be challenging to definitively confirm  at the LHC \cite{Duhrssen:2004cv}, but would be easily measured at a linear collider. For a recent overview, see \cite{Peskin:2012we}.  While deviations in the Higgs couplings motivate this model, potentially just as interesting are the direct searches for the new states introduced here.  The largest modifications to the $\gamma \gamma$ rate are only possible for for light leptons with masses, very likely accessible soon at the LHC.

A similar Lagrangian was considered in \cite{Phalen:2009xw} in an attempt to explain the PAMELA positron excess.  There, a single vectorlike lepton was introduced (rather than both $E$ and $L$).  (In that case, all corrections are proportional to $y_{\tau}^0$, and will not be numerically significant.)  However, just as in that case, if the Dark Matter (here left unspecified) couples  to the vectorlike leptons introduced here, it could lead to relatively leptophilic annihilation channels, with important implications for indirect detection.

The discovery of the Higgs boson is a watershed moment in particle physics.  If new physics exists at the weak scale, it may leave imprints on the measurable properties of the Higgs boson.  What is remarkable about a 125 GeV Higgs is that, with many observables, its properties are so sensitive to many scenarios of BSM physics. The presence of new vectorlike fermions is intriguing, because a complete generation can influence independently $\Gamma_{h \rightarrow \gamma \gamma}$, $gg$-fusion, $\Gamma_{h \rightarrow \tau \bar{\tau}}$ and $\Gamma_{h \rightarrow b \bar{b}}$. Should future data hold up an anomaly in these quantities, it may point to new matter just around the corner.

{\bf Note Added:}
While this paper was nearing completion, some works \cite{Joglekar:2012hb,ArkaniHamed:2012kq,Almeida:2012bq}) appeared that discussed the enhancement to $\gamma \gamma$ due to vectorlike fermions.  Also, \cite{Pospelov} noted the possibility of suppressing the $\tau$ Yukawa via a mixing with vectorlike leptons in a similar model with an additional singlet scalar.

\begin{acknowledgments}
The work of A.P. was supported in part by NSF CAREER Grant NSF-PHY-0743315  and by DOE Office of Science under Grant DE-SC0007859 . The work of J.K. was supported by DOE Office of Science under Grant DE-SC0007859. NW is supported by NSF grant \#0947827. AP acknowledges the Aspen Center for Physics where some of this work was completed. 
\end{acknowledgments}

\appendix
\section{Higgs Loop Functions}
For completeness, we reproduce the loop functions for the generation of the $h \rightarrow \gamma \gamma$ process used in Sec.~\ref{sec:hgg}.
\begin{eqnarray}
A_{1/2} (\tau) &=& 2 \tau^2 \left( \frac{1}{\tau} + \left(\frac{1}{\tau} -1\right) f\left(\frac{1}{\tau}\right)\right),\\
A_{1}(\tau)&=& -\tau^2 \left( \frac{2}{\tau^{2}} + \frac{3}{\tau} + 3\left(\frac{2}{\tau}-1\right) f\left(\frac{1}{\tau}\right)\right),\\
f(x) &=& \rm{Arcsin}^{2} \sqrt{x}.
\end{eqnarray}
The form of $f(x)$ is valid for $m_h < 2 m_i$, where $m_i$ is the mass of the particle in the loop, as for the particles considered in this paper.

\section{Functions for the T parameter}
\label{app:rhoparameter}
Neglecting mixing with the Standard Model particles (which is in any case constrained to be small), we find a contribution to the isospin breaking $T$ parameter as \cite{Lavoura:1992np}.  In this limit, the mixing of the charged vectorlike leptons is described by the $2 \times 2$ mass matrix
\begin{equation}
-{\mathcal L}_{\text{mass}} =  \left (\begin{array}{cc} 
E^{c} & \bar{L}^+
\end{array}
\right) 
\left(
\begin{array}{cc}
M_{E} & y_{LE} v\\
 \bar{y}_{LE} v & M_{L}
\end{array}
\right)
\left(
\begin{array}{c}
\bar{E}^{c}\\
L^-
\end{array}
\right) + \text{h.c.},
\end{equation}
which can be diagonalized to yield mass eigenstates
\begin{equation}
\left(\begin{array}{c} \ell_1 \\ \ell_2 \end{array}\right) = V \left(\begin{array}{c} \bar{E}^c \\ L^- \end{array}\right) \text{ and } \left(\begin{array}{c} \bar{\ell}_1 \\ \bar{\ell}_2 \end{array}\right) = U^\ast \left(\begin{array}{c} E^c \\ \bar{L}^+ \end{array}\right), 
\end{equation}
where $\mathcal{M}_D = \text{diag}(m_1,m_2) = U \mathcal{M} V^\dagger$.  The spectrum also contains of a vectorlike neutral state with mass $M_L$.  Modifying the expression from \cite{Lavoura:1992np}, we find the contribution to the $T$ parameter
\begin{eqnarray}
\Delta T & = & \frac{N_c}{16 \pi s_W^2 c_W^2} \left\{\sum_i \Big((\abs{V_{i2}}^2 + \abs{U_{i2}}^2) \theta_+\left(y_i, y_L\right) + 2 \text{Re}(V_{i2} U^\ast_{i2}) \theta_-\left(y_i, y_L\right)\Big) \right. \nonumber \\ & & - \left. \vphantom{\sum_i} \Big(\left(\abs{V_{12} V_{22}}^2 + \abs{U_{12} U_{22}}^2\right) \theta_+\left(y_1, y_2\right) + 2 \text{Re}(V_{12} V_{22}^\ast U_{12}^\ast U_{22}) ) \theta_-\left(y_1, y_2\right)\Big) \right\}
\end{eqnarray}
where $y_i = m_i^2/m_Z^2$, $N_c$ is the number of colors and
\begin{eqnarray}
\theta_+(y_1,y_2) & = & y_1 + y_2 - \frac{2 y_1 y_2}{y_1 - y_2} \log\frac{y_1}{y_2} \\
\theta_-(y_1,y_2) & = & 2 \sqrt{y_1 y_2} \left(\frac{y_1 + y_2}{y_1 - y_2} \log \frac{y_1}{y_2} - 2\right).
\end{eqnarray}
Completely analogous expressions hold for the quark model.

\bibliography{Tauphobic}
\end{document}